\documentclass[aps,prb,reprint,twocolumn,superscriptaddress,citeautoscript,amsmath,amssymb,floatfix]{revtex4-2}
\usepackage{graphicx}
\usepackage{times}
\usepackage{bm}
\usepackage{color}
\usepackage{gensymb}
\usepackage{dcolumn}
\usepackage{xfrac}
\usepackage{booktabs}

\usepackage{adjustbox}
\newcolumntype{P}[1]{>{\centering\arraybackslash}p{#1}}
\usepackage{lipsum}

\usepackage{textcomp}
\newcommand{\tildeapprox}{\raisebox{0.5ex}{\texttildelow}}

\newcommand{\LaNi}[0]{$\text{La}_{2}\text{Ni}_{7}$}

\newcommand{\TNCc}[0]{$T_{3}$}
\newcommand{\TNICab}[0]{$T_{2}$}
\newcommand{\TNICc}[0]{$T_{1}$}
\newcommand{\Cc}[0]{$\text{A-phase}$}
\newcommand{\ICab}[0]{$\text{B-phase}$}
\newcommand{\ICc}[0]{$\text{C-phase}$}

\newcolumntype{d}[1]{D{.}{.}{#1}}

\begin{document}
\title{Weak itinerant magnetic phases of \LaNi{}}

\author{J.~M.~Wilde}
\affiliation{Ames Laboratory, U.S. DOE, Iowa State University, Ames, Iowa 50011, USA}
\affiliation{Department of Physics and Astronomy, Iowa State University, Ames, Iowa 50011, USA}

\author{A. Sapkota}
\affiliation{Ames Laboratory, U.S. DOE, Iowa State University, Ames, Iowa 50011, USA}
\affiliation{Department of Physics and Astronomy, Iowa State University, Ames, Iowa 50011, USA}

\author{W. Tian}
\affiliation{Neutron Scattering Division, Oak Ridge National Laboratory, Oak Ridge, Tennessee 37831, USA}

\author{S.~L.~Bud'ko}
\affiliation{Ames Laboratory, U.S. DOE, Iowa State University, Ames, Iowa 50011, USA}
\affiliation{Department of Physics and Astronomy, Iowa State University, Ames, Iowa 50011, USA}

\author{R.~A.~Ribeiro}
\affiliation{Ames Laboratory, U.S. DOE, Iowa State University, Ames, Iowa 50011, USA}
\affiliation{Department of Physics and Astronomy, Iowa State University, Ames, Iowa 50011, USA}

\author{A. Kreyssig}
\affiliation{Ames Laboratory, U.S. DOE, Iowa State University, Ames, Iowa 50011, USA}
\affiliation{Department of Physics and Astronomy, Iowa State University, Ames, Iowa 50011, USA}
\affiliation{Institute for Experimental Physics 4, Ruhr-Universit\"{a}t Bochum, 44801 Bochum, Germany}

\author{P.~C.~Canfield}
\affiliation{Ames Laboratory, U.S. DOE, Iowa State University, Ames, Iowa 50011, USA}
\affiliation{Department of Physics and Astronomy, Iowa State University, Ames, Iowa 50011, USA}

\date{\today}

\begin{abstract}
\LaNi{} is an intermetallic compound that is thought to have itinerant magnetism with small moment (\tildeapprox0.15~$\mu_{\text{B}}/\text{Ni}$) ordering below 65~K. A recent study of single crystal samples by Ribeiro $et.~al.$ [Phys. Rev. B \textbf{105}, 014412 (2022)] determined detailed anisotropic $H$-$T$ phase diagrams and revealed three zero-field magnetic phase transitions at \TNICc{}\tildeapprox61.0~K, \TNICab{}\tildeapprox56.5~K, and \TNCc{}\tildeapprox42~K. In that study only the highest temperature phase is shown to have a clear ferromagnetic component. Here we present a single crystal neutron diffraction study determining the propagation vector and magnetic moment direction of the three magnetically ordered phases, two incommensurate and one commensurate, as a function of temperature. The higher temperature phases have similar, incommensurate propagation vectors, but with different ordered moment directions. At lower temperatures the magnetic order becomes commensurate with magnetic moments along the $\bm{c}$ direction as part of a first-order magnetic phase transition. We find that the low-temperature commensurate magnetic order is consistent with a proposal from earlier DFT calculations.
\end{abstract}

\maketitle

\section{Introduction}

One route to study interesting phenomena such as superconductivity or non-Fermi liquid behavior has been to focus on the role of fluctuations associated with a quantum critical point (QCP) where a continuous phase transition occurs at zero temperature ($T$)\cite{stewart2001non,canfield2016preserved,canfield2020new}. One way to accomplish this is to suppress existing magnetic order to $T=0$ using external control parameters such as pressure and/or chemical substitution \cite{pfleiderer2001non,levy2007acute,uemura2007phase,huy2007superconductivity,cheng2015pressure,brando2016metallic,ran2019nearly,belitz1997nonanalytic,canfield2016preserved,canfield2020new}. There is a large body of experimental evidence showing a QCP can be accessed for antiferromagnetic (AFM) transitions such as in iron-based superconductors \cite{shibauchi2014quantum} and in heavy-fermion metals \cite{gegenwart2008quantum}. Weak itinerant magnets with their large spin fluctuations are interesting to study since they can often be driven into novel phases around a QCP\cite{moriya1992antiferromagnetic, ishigaki1998theory, moriya2003antiferromagnetic}. For weak itinerant ferromagnets (FM) there is unfortunately a growing body of evidence that a QCP is often entirely avoided such as for $\text{La}\text{Cr}\text{Ge}_{3}$\cite{taufour2016ferromagnetic,kaluarachchi2017tricritical,gati2021formation}, and $\text{La}_5\text{Co}_2\text{Ge}_{3}$\cite{xiang2021avoided}. Of particular interest are weak itinerant AFM such as $\text{Ti}\text{Be}_{2}$\cite{matthias1978itinerant,povzner1995spin,torun2016origin}, $\text{Ti}\text{Au}$\cite{svanidze2015itinerant,goh2016mechanism,goh2017competing,mathew2019probing}, 122-type cobalt arsenides\cite{sangeetha2019non,wilde2019helical,li2019flat,li2019antiferromagnetic}, $\text{Mn}\text{Si}$\cite{kadowaki1982magnetization,stishov2011itinerant}, and $\text{Lu}\text{Fe}_2\text{Ge}_2$\cite{avila2004anisotropic,fujiwara2007pressure} . 

\begin{figure}[htbp]
    \centering
    \includegraphics[width=\columnwidth]{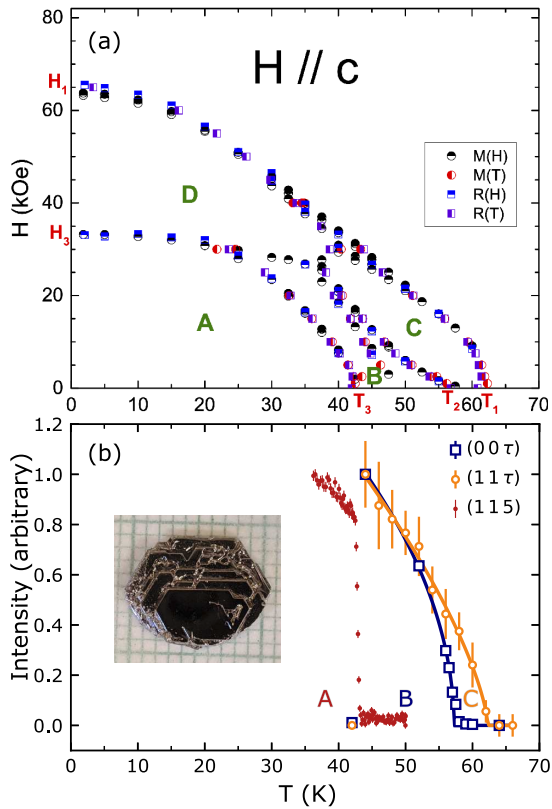}
    \caption{(a) $H$-$T$ phase diagram when $\bm{\text{H}}\parallel\bm{c}$ reproduced from data shown in Ribeiro $et.~al.$ \cite{ribeiro2022small}. (b) Temperature dependence of $(0\,0\,\tau)$, $(1\,1\,\tau)$, and $(1\,1\,5)$ AFM Bragg peaks in blue, orange and red, respectively. For $(0\,0\,\tau)$ and $(1\,1\,\tau)$ the intensity used is the integrated intensity and for $(1\,1\,5)$ intensity was measured on top of the Bragg peak. The highest measured intensity for each peak is normalized to one with solid lines are a power law fit for $(0\,0\,\tau)$ and $(1\,1\,\tau)$. Inset shows single crystal of \LaNi{} used in neutron diffraction measurement.}
    \label{Order_params}
\end{figure}

 \LaNi{} is an example of a small-moment itinerant magnetic system which has had its magnetic properties studied for many years \cite{buschow1983magnetic,parker1983magnetic,gottwick1985transport,tazuke1993magnetism,tazuke1997magnetic,fukase1999successive,fukase2000itinerant,tazuke2004metamagnetic,crivello2020relation,ribeiro2022small}, but until recently only in its polycrystalline form. Recently, \LaNi{} has been successfully synthesised as large single crystals \cite{ribeiro2022small}. Magnetization, specific heat, and electric transport measurements on these single crystals were used to assemble anisotropic $H$-$T$ phase diagrams [see Fig.~\ref{Order_params}(a)] for the applied field parallel to the crystallographic $c$-axis). For zero applied field, three distinct phase transitions at \TNICc{}\tildeapprox61.0~K, \TNICab{}\tildeapprox56.5~K, and \TNCc{}\tildeapprox42~K can be seen delineating three magnetically ordered regions that we refer to as \Cc{} for $T<$~\TNCc{}, \ICab{} for \TNCc{}~$<T<$~\TNICab{}, and \ICc{} for \TNICab{}~$<T<$~\TNICc{} \cite{ribeiro2022small}. Previous attempts to measure the possible antiferromagnetic order using neutron powder diffraction failed to detect magnetic Bragg peaks at low temperature and set an upper limit of 0.03~$\mu_{\text{B}}/\text{Ni}$ for the size of the ordered moments\cite{tazuke1997magnetic}. 
 
\begin{figure}[htbp]
    \centering
    \includegraphics[width=\columnwidth]{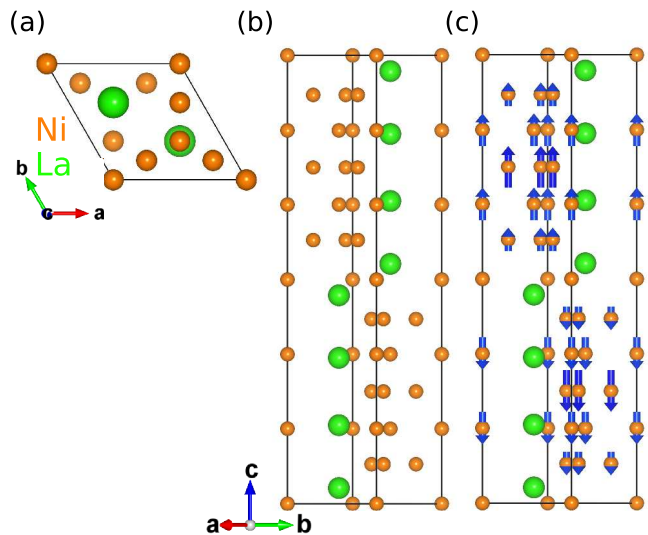}
    \caption{Chemical and proposed magnetic structure of the \Cc{} of \LaNi{} generated by \textsc{vesta} \cite{Momma_2011}. (a) ab-plane of hexagonal chemical structure. (b) Chemical structure of \LaNi{} (c) Magnetic structure for the \Cc{} with magnetic space group $P 6_3^{\prime} / mm^{\prime}c$.}
    \label{Structure}
\end{figure}

 Here we report the results of a single crystal neutron diffraction study of the three AFM phases for zero applied field in \LaNi{}. The results of this study are summarized in Table~\ref{Tab:summary_phases}. We found the \Cc{} to be commensurate, and the \ICab{} and \ICc{} to be incommensurate below 42.8(5), 57.2(5), and 62.3(7), respectively. We find that the transition temperatures \TNICc{}, \TNICab{}, and \TNCc{} is consistent with the $H$-$T$ phase diagram determined by Ribeiro $et.~al.$ \cite{ribeiro2022small}. We demonstrate that the \ICc{} and \ICab{} have the same temperature dependent propagation vector ($\tau$), but with different magnetic moment directions. The change in magnetic moment direction coincides with the disappearance of the net ferromagnetic component in the magnetic ordering of the \ICc{} upon cooling into the \ICab{} observed by Ribeiro $et.~al.$ \cite{ribeiro2022small}. Upon further cooling, the magnetic structure changes from incommensurate to commensurate with a first-order transition. In addition, we find the intensity distribution of several magnetic Bragg peaks in the \Cc{} are consistent with a triangular wave modulated SDW with moments along the $\bm{c}$ direction shown in Fig.~\ref{Structure}(c), and as theoretically predicted recently \cite{crivello2020relation}.
 
\section{Experimental Methods}

Single crystals of \LaNi{} were grown out of a La-rich (relative to \LaNi{}) binary, high-temperature melt. Elemental La (Ames Laboratory, 99.99+\%{} pure) and Ni (Alpha, 99.9+\%{}) were weighed out in a $\text{La}_{33}\text{Ni}_{67}$ atomic ratio and placed into a tantalum crucible which was sealed with solid caps on each end and a fritted cap in the middle to act as a frit or filter for decanting \cite{canfield2001high, ribeiro2022small, canfield2020new}. The assembled Ta crucible was then itself sealed into an amorphous silica tube with silica wool above and below for cushioning. This growth ampoule was then placed in a resistive box furnace. The furnace was then heated to 1150~\degree C for 10 hours, held at 1150~\degree C for 10 hours, cooled to 1020~\degree C over 4 hours, and then very slowly cooled to 820~\degree C over 300 hours, at which point the growth ampoule was removed and decanted in a centrifuge to separate the \LaNi{} single crystals from the residual liquid \cite{canfield2020new}. Crystals grew as well faceted plates with clear hexagonal morphology as seen in the inset of Fig.~\ref{Order_params}(b). The surfaces of the crystal are orthogonal to the [0\,0\,1] direction, and the edges of the facets are along the [1\,1\,0] direction.

Neutron diffraction measurements were performed on a 696 mg single crystal using the HB-1A FIE-TAX triple-axis spectrometer at the High Flux Isotope Reactor, Oak Ridge National Laboratory. FIE-TAX operates at a fixed incident energy of 14.7 meV using two pyrolytic graphite (PG) monochromators. Two PG filters are place before and after the second monochromator to significantly reduce higher harmonics within the incident beam. The beam collimators placed before the monochromator, between the monochromator and sample, between the sample and analyzer, and between the analyzer and detector were $40^{\prime}-40^{\prime}-\text{sample}-40^{\prime}-80^{\prime}$, respectively. Samples were sealed in an Al can containing He exchange gas which was then attached to the cold head of a closed-cycle He refrigerator. Scattering data are described using reciprocal lattice units of $H$, $K$, and $L$ for the hexagonal unit cell of \LaNi. The sample we used is shown in the inset of Fig.~\ref{Order_params}(b) which was aligned with the $(H~H~L)$ reciprocal-lattice plane coincident with the spectrometer's scattering plane.
\begin{figure}[htbp]
    \centering
    \includegraphics[width=\columnwidth]{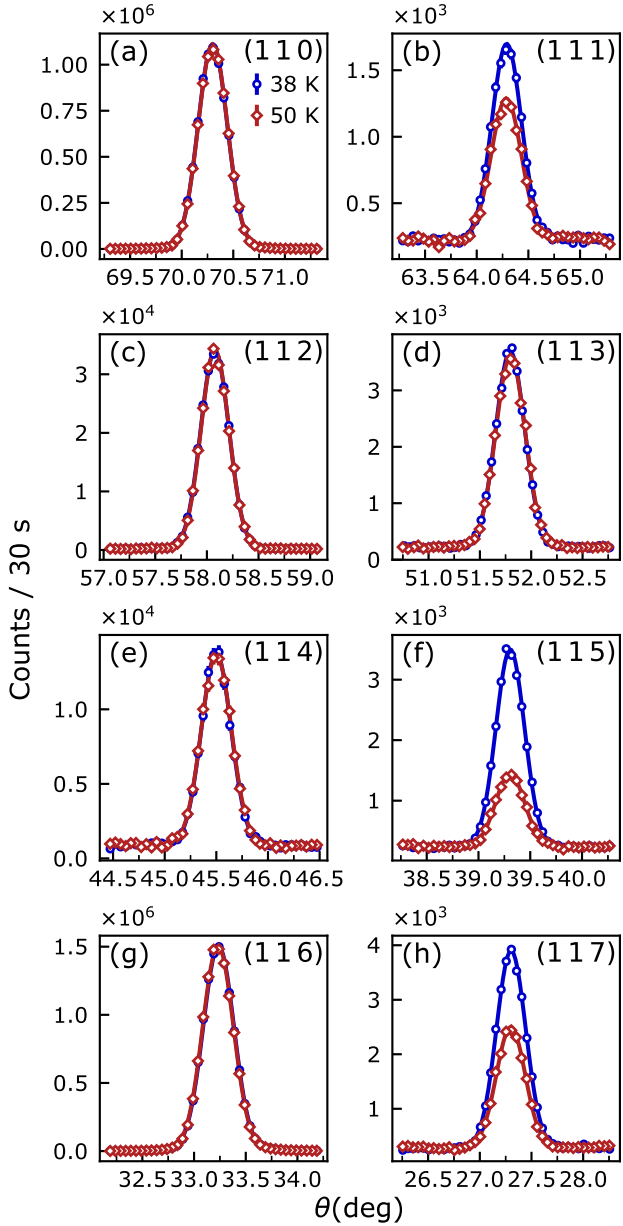}
    \caption{Rocking scans ($\theta$) of Bragg peaks along $(1\,1\,L)$ for
    \LaNi. Red and blue indicate measurements taken above and below \TNCc{} which corresponds to 50~K (\ICab{}) and 38~K (\Cc{}), respectively. The data are normalized to 30 mcu which corresponds to 30 seconds of counting time. Fits to the data were made using a Gaussian line-shape.}
    \label{C_Rocking_11L}
\end{figure}

\begin{figure}[htbp]
    \centering
    \includegraphics[width=\columnwidth]{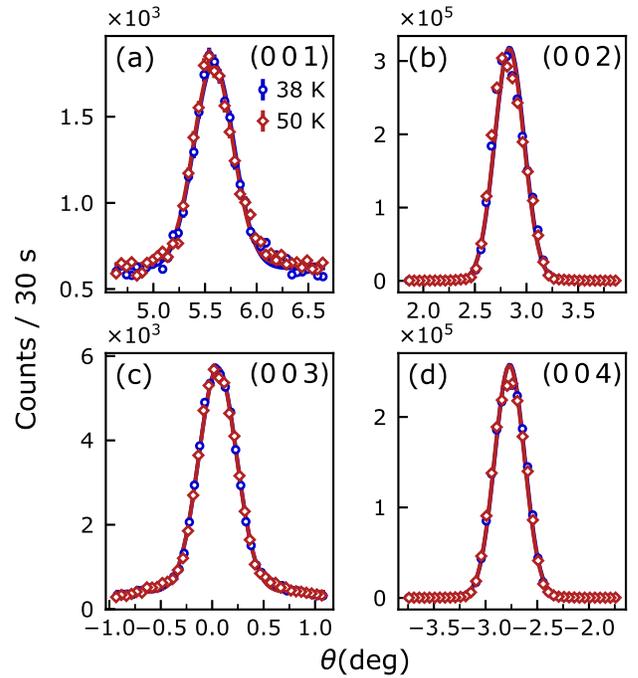}
    \caption{Rocking scans ($\theta$) of Bragg peaks along $(0\,0\,L)$ for
    \LaNi. Red and blue indicate measurements taken above and below \TNCc{} which corresponds to 50~K (\ICab{}) and 38~K (\Cc{}), respectively. The data are shown with a logarithmic scale for the count that are normalized to 30 mcu which corresponds to 30 seconds of counting time. Fits to the data were made using a Gaussian line-shape.}
    \label{C_Rocking_00L}
\end{figure}

\section{Results and Discussion}

\LaNi{} crystallizes in the hexagonal $\text{Ce}_{2}\text{Ni}_{7}$ structure type with space group $P 6_3 / mmc$ as shown in Fig.~\ref{Structure}(a) and (b) with lattice parameters $a=b=5.06352(11)~\text{\AA}$ and $c=24.6908(8)~\text{\AA}$ at room temperature\cite{ribeiro2022small}. Furthermore, the structure consists of 2 separate La sites at the 4f Wyckoff position, and 5 sites for Ni located at the 12k, 6h, 4f, 4e, and 2a Wyckoff positions. Reflection conditions for the nuclear Bragg peaks are as follows: $(0\,0\,L)$ with $L=2n$, and $(H\,H\,L)$ with $L=2n$, where $n$ is an integer. 
 
\begin{figure}[htbp]
    \centering
    \includegraphics[width=\columnwidth]{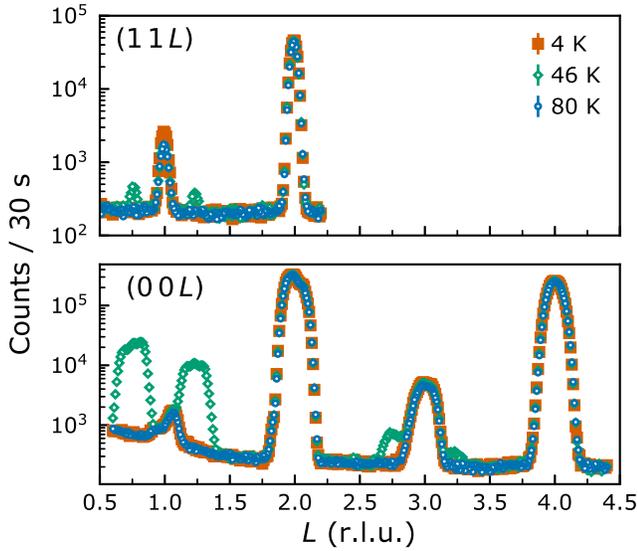}
    \caption{Scans along the $L$ direction taken within the \Cc{}, \ICab{}, and PM-phase at 4~K (\Cc{}), 46~K (\ICab{}) and 80~K (PM-phase) shown as green squares, blue diamonds, and red circles, respectively. The data is normalized to 30 mcu which corresponds to 30 s of counting time.}
    \label{00L_Longscan}
\end{figure}

 The \LaNi{} structure can be described as a structure made up of blocks of $\text{La}_2\text{Ni}_4$ and $\text{La}\text{Ni}_5$, respectively. For the \LaNi{} structure these form block layers of [$\text{La}_2\text{Ni}_4 +2\text{La}\text{Ni}_5$] which can have an AB or ABC stacking arrangement resulting in a hexagonal (2H) or rhombohedral (3R) structure, respectively, and as shown in detail in several studies\cite{di2000characterization,crivello2020relation}. \LaNi{} used in this study and several others crystallizes with the 2H structure, but with the presence of both intra-block and inter-block stacking faults. The stacking faults can break the general reflection conditions in our system, resulting in intensity at forbidden peaks with $L=odd$ at all temperatures as seen in both rocking ($\theta$) scans and scans along the $L$ direction of $(0\,0\,L)$ and $(1\,1\,L)$ peaks, as shown in Figs.~\ref{C_Rocking_11L},  \ref{C_Rocking_00L}, and \ref{00L_Longscan}. This is consistent with previous electron diffraction studies showing forbidden peaks at $(H\,H\,L)$ with $L=odd$ due to the presence of block stacking faults \cite{di2000characterization}.

Figures~\ref{C_Rocking_11L} and \ref{C_Rocking_00L} show rocking ($\theta$) scans at temperatures within the \Cc{} (38~K) and the \ICab{} (50~K) for the $(1\,1\,L)$ and $(0\,0\,L)$ Bragg peaks. Below \TNCc{}, additional magnetic intensity occurs at Bragg peak positions commensurate to the unit cell $(1\,1\,1)$, $(1\,1\,5)$, and $(1\,1\,7)$ as shown by Fig.~\ref{C_Rocking_11L}(b), \ref{C_Rocking_11L}(f), and \ref{C_Rocking_11L}(h), respectively. The magnetic Bragg peak positions correspond to a propagation vector $\bm{\tau}=(0\,0\,1)$. Furthermore, the magnetic Bragg peaks do not correspond to any simple AFM structure, since the magnetic Bragg peak intensity does not monotonically decrease with increasing scattering vector, $q$, as would be expected for the magnetic form factor of $\text{Ni}^{2+}$. For example, additional $(1\,1\,3)$ magnetic Bragg peak intensity is conspicuously absent as shown in Fig.~\ref{C_Rocking_11L}(d), but clearly observed for the $(1\,1\,5)$ magnetic Bragg peak as shown in Fig.~\ref{C_Rocking_11L}(f). AFM order within the \Cc{} is long-range since the magnetic Bragg peaks have similar full width at half maximum (FWHM) to their respective nearby nuclear Bragg peaks. No change in the shape or intensity of Bragg peaks are observed at positions $(1\,1\,L)$, $L\,{=}\,even$ as seen in Fig.~\ref{C_Rocking_11L}. In addition, no additional intensity is observed at any $(0\,0\,L)$ position as seen in Fig.~\ref{C_Rocking_00L} which suggests that the magnetic structure of the commensurate magnetic phase has ordered moments $\bm{\mu} \parallel \bm{c}$. All observed Bragg peaks were fit using one Gaussian in order to determine their integrated intensity and FWHM.

\begin{figure}[htbp]
    \centering
    \includegraphics[width=\columnwidth]{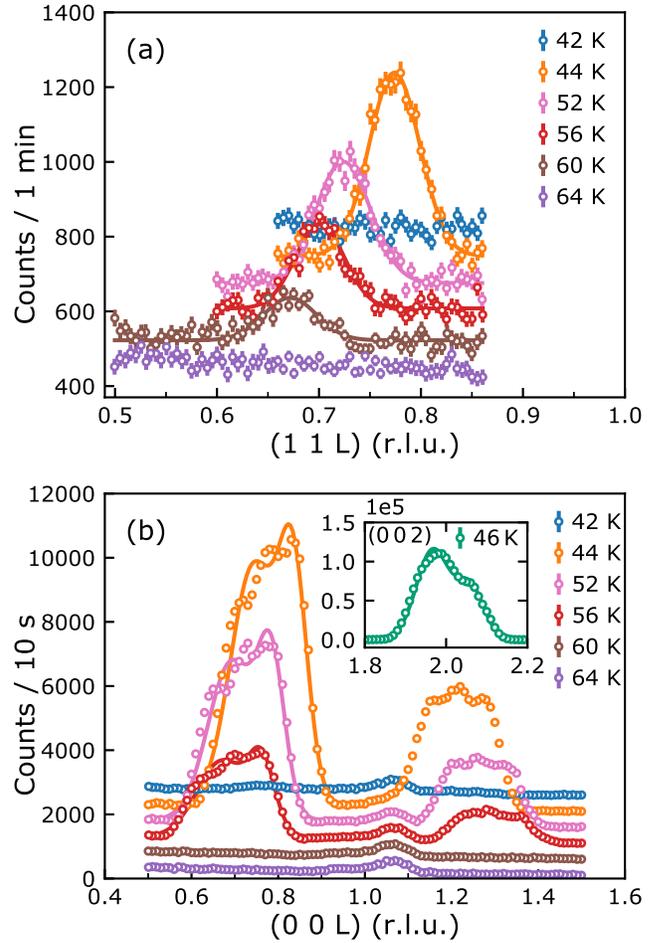}
    \caption{Relative scans along $L$ for AFM Bragg peak (a) $(1\,1\,\tau)$ and (b) $(0\,0\,\tau)$ and $(0\,0\,2{-}\tau)$. The data for (a) was taken with 60 mcu which corresponds to 1 minute, and the data for (b) was taken with 10 mcu which corresponds to 10 s of counting time. Inset for (b) shows a relative scan along $L$ for the nuclear Bragg peak $(0\,0\,2)$ normalized to 10 mcu at 46~K. Fits to the data were made made using a Gaussian line shape and a two-Gaussian line shape for (a) and (b), respectively. Data are offset for clarity.}
    \label{IC_Magnetic-L}
\end{figure}

\begin{figure}[htbp]
    \centering
    \includegraphics[width=\columnwidth]{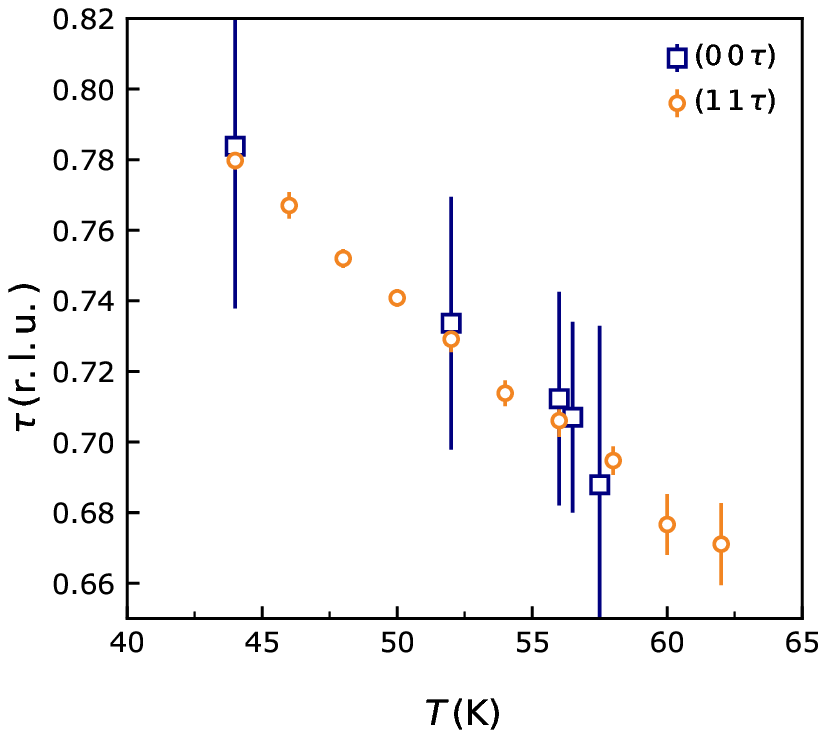}
    \caption{Temperature dependence of $\tau$ determined by the position of magnetic Bragg peaks $(0\,0\,\tau)$ and $(1\,1\,L)$ in blue and orange, respectively.}
    \label{TauvsT}
\end{figure}

Figure~\ref{00L_Longscan} shows scans along $L$ taken at 4, 46, and 80~K, which are in the \Cc{}, \ICab{}, and paramagnetic (PM) phase, respectively. Additional magnetic intensity is apparent at 46~K with $\tau\approx0.7$, and is absent at 4~K. At selected temperatures scans along $L$ of magnetic Bragg peaks $(0\,0\,\tau)$, $(1\,1\,\tau)$, and $(0\,0\,2{-}\tau)$ are shown in Fig.~\ref{IC_Magnetic-L}. The Bragg peak $(1\,1\,L)$ at each temperature was fit with a single Gaussian and a linear background in order to determine the magnetic peak intensity and $\tau$ as a function of temperature as shown for 44, 52, 56, and 60~K in Fig.~\ref{IC_Magnetic-L}(a). Since the longitudinal scans of $(0\,0\,\tau)$ and $(0\,0\,2{-}\tau)$ are very structured and the peaks are very intense, the intensity at each temperature was determined by numerical integration. To determine $\tau$ for the the $(0\,0\,\tau)$ Bragg peaks the longitudinal scan of the $(0\,0\,2)$ Bragg peak at 46~K was first fit with a two Gaussian function with a constant background as shown in the inset of Fig.~\ref{IC_Magnetic-L}(b). The relative position of each Gaussian was fixed and used to fit longitudinal scans of $(0\,0\,\tau)$. The position of each Gaussian was then averaged to obtain $\tau$ for the $(0\,0\,\tau)$ peak. The two Gaussian fit of $(0\,0\,\tau)$ does not capture all the details of the structured peak, but does reproduce the width of each Bragg peak reasonably well as shown in Fig.~\ref{IC_Magnetic-L}(b) at 44, 52, and 56~K.

Several observations can be made qualitatively at this point based on data shown in Fig.~\ref{00L_Longscan} and \ref{IC_Magnetic-L}. In Fig.~\ref{00L_Longscan}, magnetic and nuclear Bragg peaks are structured in a similar manner indicating that the structure is not related to details of the magnetic and chemical order, but to extrinsic effects, e.g. from the combined effect of sample and instrument profile, due to the high-quality of the sample, it's large shape, and slab-like geometry. In Fig.~\ref{IC_Magnetic-L} the intensity of the magnetic Bragg peaks $(1\,1\,\tau)$, $(1\,1\,2{-}\tau)$, and $(0\,0\,\tau)$ increases steadily with decreasing temperature until they abruptly vanish at \tildeapprox42~K. This corresponds to the first-order transition between \ICab{} and \Cc{} which occurs at \TNCc{}. Furthermore, $\tau$ for each peak clearly increases with decreasing temperature as shown in Fig.~\ref{TauvsT}. The intensity of the $(1\,1\,\tau)$ Bragg peak is much smaller than the $(0\,0\,\tau)$ and $(0\,0\,2{-}\tau)$ Bragg peaks within the \ICab{}. The $(0\,0\,\tau)$ magnetic Bragg peak indicates a magnetic moment component $\bm{\mu} \perp \bm{c}$.

\begin{table*}
	\caption{ \label{Tab:summary_phases}  Summary of the properties of AFM phases of \LaNi{} determined by single crystal neutron diffraction.} 
	\begin{ruledtabular}
		\begin{tabular}{lcccc}		
		& Phase A & Phase B & Phase C \\
		    \noalign{\vskip 1mm}
			\hline
			\noalign{\vskip 1mm}
			 Temperature & $T<42.8(5)\,\text{K}$ & $42.8(5)<T<57.2(5)\,\text{K}$& $57.2(5)<T<62.3(7)\,\text{K}$ \\
			 \noalign{\vskip 1mm}
			 Propagation vector & $(0\,0\,1)$ & $(0\,0\,\tau)$, $\tau\tildeapprox{}0.78\rightarrow0.69$ & $(0\,0\,\tau)$, $\tau\tildeapprox{}0.69\rightarrow0.67$ \\
			 \noalign{\vskip 1mm}
			 Magnetic moment direction & $\bm{\mu}\parallel\bm{c}$ & $\bm{\mu}\parallel\bm{c}$ and $\bm{\mu}\perp\bm{c}$ components & $\bm{\mu}\perp\bm{c}$ \\
			
		\end{tabular}
	\end{ruledtabular}
\end{table*}

Figure \ref{Order_params}(b) shows the temperature dependence of the integrated intensity of the $(0\,0\,\tau)$ and $(1\,1\,\tau)$ Bragg peak positions, and the intensity sitting on top of the $(1\,1\,5)$ Bragg peak position. The intensity of the $(1\,1\,5)$ Bragg peak was measured this way since no significant changes in the lattice parameters were detected across \TNCc{}. The temperature dependence of each second-order transition in \ref{Order_params}(b) are shown with a fit to a power law with the results \TNICab{}~=~57.2(5)~K and $\beta$~=~0.25(0.09) for $(0\,0\,\tau)$, and \TNICc{}~=~62.3(7)~K and $\beta$~=~0.35(0.04) for $(1\,1\,\tau)$. The AFM ordering of Bragg peak $(1\,1\,5)$ and magnetic Bragg peaks with propagation vector $\bm{\tau}=(0\,0\,1)$ vanishing is consistent with a first-order transition. The transition temperature, \TNCc{}~=~42.8(5)~K, was determined by the minimum of the numerical derivative of the intensity of the $(1\,1\,5)$ Bragg peak. Neutron diffraction is directly sensitive to the order parameter of magnetic transitions, because the intensity of magnetic Bragg peaks is proportional to the square of the magnetic moment component $\bm{\mu} \perp \bm{q}$. Upon cooling below \TNICc{}~=~62.3(7)~K the \ICc{} develops with a component of $\bm{\mu}\parallel\bm{c}$, as shown by the appearance of the magnetic Bragg peak $(1\,1\,\tau)$ and the absence of the Bragg peak $(0\,0\,\tau)$ in Fig.~\ref{Order_params}(b). The appearance of the Bragg peak $(0\,0\,\tau)$ below \TNICab{}~=~57.2(5)~K indicates that the magnetic structure develops an additional component of $\bm{\mu}\perp\bm{c}$ in the \ICab{}. Below \TNCc{}~=~42.8(5)~K the \Cc{} has a unit cell identical to the chemical unit cell, and the magnetic Bragg peak $(1\,1\,5)$ with no $(1\,1\,L)$, $L\,{=}\,odd$ magnetic Bragg peak indicates the magnetic moment is likely $\bm{\mu}\parallel\bm{c}$. 

Figure~\ref{TauvsT} shows in detail the temperature dependence of the propagation vector $\tau$ for the Bragg peaks $(0\,0\,\tau)$ and $(1\,1\,\tau)$. Remarkably, $\tau$ has a linear dependence with temperature which increases with decreasing temperature from a minimum value of ~0.68 at 62~K as shown in Fig.~\ref{TauvsT}. There is good agreement of the $\tau$ at each temperature determined from the magnetic Bragg peaks $(0\,0\,\tau)$ and $(1\,1\,\tau)$. This indicates that the phase transition at \TNICab{} is related to the development of the magnetic moment component $\bm{\mu}\perp\bm{c}$ in phase B in addition to the already present component $\bm{\mu}\parallel\bm{c}$ in phase C, e.g. the AFM moments $\bm{\mu}\perp\bm{c}$ in phase C tilt away from the $c$-axis in phase B below \TNICab{}.

\subsection{Discussion}

\LaNi{} has five Ni sites and so determining a unique solution to the magnetic structure is not possible without access to a large number of magnetic and nuclear Bragg peaks. As stated previously the phase transition at \TNCc{} is first order, but since there is no observed structural transition below room temperature we can use group-subgroup relations to explore possible magnetic space groups (MSG). Using the Bilbao Crystallographic Server\cite{aroyo2006bilbao}, we found eight maximal isomorphic subgroups of $P 6_3 / mmc$ with propagation vector $(0\,0\,1)$ and every Ni site as a possible magnetic site. These 8 MSG are: $P 6_3 / m^{\prime}m^{\prime}c^{\prime}$, $P 6_3 / mm^{\prime}c^{\prime}$, $P 6_3^{\prime} / m^{\prime}mc^{\prime}$, $P 6_3^{\prime} / m^{\prime}m^{\prime}c$, $P 6_3^{\prime} / mmc^{\prime}$, $P 6_3^{\prime} / mm^{\prime}c$, $P 6_3 / m^{\prime}mc$, and $P 6_3 / mmc$. To agree with our observations we also require a magnetic moment component of $\bm{\mu} \parallel \bm{c}$, so the possible MSG can be reduced to 4: $P 6_3 / m^{\prime}m^{\prime}c^{\prime}$, $P 6_3 / mm^{\prime}c^{\prime}$, $P 6_3^{\prime} / m^{\prime}m^{\prime}c$, and $P 6_3^{\prime} / mm^{\prime}c$. 

Recent DFT calculations by Crivello and Paul-Boncour have predicted that for AFM order \LaNi{} would preferentially order along $\bm{c}$ as a triangle-wave modulated spin-density wave as shown in Fig.~\ref{Structure}(c) \cite{crivello2020relation}. This magnetic structure is consistent with the possible MSG $P 6_3^{\prime} / mm^{\prime}c$. In addition, from Fig.~\ref{C_Rocking_11L} we see that the sequence of intensities for Bragg peaks $(1\,1\,1)$, $(1\,1\,3)$, $(1\,1\,5)$, and $(1\,1\,7)$ do not only decrease as a function of $q$, but are instead consistent with the structure shown in Fig.~\ref{Structure}(c) with average magnetic moment given by $\mu_{\text{Ni}}=0.19(4) \mu_{\text{B}}/\text{Ni}$ in the \Cc{}. To obtain the average magnetic moment, the measured integrated intensities of AFM Bragg peaks are compared to intensities calculated using FULLPROF~\cite{rodriguez1993fullprof} for a specific spin motif consistent with a magnetic subgroup. The calculated and experimental values are compared using a scale factor determined by the intensities of the nuclear Bragg peaks. This is not yet a conclusive magnetic structural refinement, since for example the more conventional sinusoidal modulated SDW with $\bm{\mu}\parallel\bm{c}$ and average magnetic moment given by $\mu_{\text{Ni}}=0.23(5) \mu_{\text{B}}/\text{Ni}$ is also consistent with our measurements. The sinusoidal modulated SDW would be part of the same MSG $P 6_3^{\prime} / mm^{\prime}c$ and so the energy difference between these spin motifs should be very small. In both cases the calculated magnetic moment is of the same order of the saturated moment of $\tildeapprox{}0.1 \mu_{\text{B}}/\text{Ni}$ measured on single crystals \cite{ribeiro2022small}. The neutron diffraction study on powder \cite{tazuke1993magnetism} likely haven't observed magnetic Bragg peaks due to the incommensurate AFM order in phases B and C, and the extremely weak AFM Bragg peaks related to it. In addition, for the \Cc{} the very weak magnetic contributions on top of Bragg peaks already existing in the high-temperature paramagnetic phase due to stacking faults within \LaNi{}.

According to Landau theory, one expects a second-order transition to decrease the symmetry of a crystal consistent with a group-subgroup relation. Since both transitions at \TNICab{} and \TNICc{} appear to be second-order we use the Bilbao Crystallographic Server \cite{aroyo2006bilbao} for the maximal isomorphic subgroups of the crystal consistent with $\bm{\tau}=(0\,0\,\tau)$ and with every Ni site as a possible magnetic site. From this we find four MSGs: $P_{\text{C}} \bar{6} c 2$, $P_{\text{C}} \bar{6} m 2$, $P_{\text{C}} \bar{3} c 1$, and $P_{\text{C}} \bar{3} m 1$. Only the MSG's $P_{\text{C}} \bar{6} c 2$ and $P_{\text{C}} \bar{3} c 1$ allow a magnetic moment component $\bm{\mu} \parallel \bm{c}$ which is likely necessary for the \ICc{}. It is possible for further symmetry to be broken for the \ICab{} below \TNICab{}, but both of the possible MSG's possible for the \ICc{} already allow magnetic moment components $\bm{\mu} \perp \bm{c}$. So we do not speculate on any lower symmetry MSG's. Unlike the \Cc{}, magnetic Bragg peak intensity for $(0\,0\,\tau)$ and $(1\,1\,\tau)$ decreases as a function of $q$ consistent with the magnetic form factor as seen in Fig.~\ref{00L_Longscan}. 

Our results show that the \Cc{}, \ICab{}, and \ICc{} are AFM within our experimental sensitivity. We are not sensitive to the relatively small FM component compared to the fully saturated \tildeapprox{}0.1~$\mu_{\text{B}}$ in the \ICc{} suggested by the $M(H)$ data shown in Ribeiro $et.~al.$ \cite{ribeiro2022small}. We are less sensitive to the proposed weak FM compared to AFM, because the additional magnetic Bragg peak intensity develops only on top of allowed nuclear Bragg peaks. It may ultimately be necessary to use magneto-optical measurements or scanning probe spectroscopy to further probe the small FM component associated with the \ICc{}. Previous powder neutron diffraction measurements may have missed the incommensurate phases since measurement were only done at 300~K and 10~K \cite{tazuke1997magnetic}. The upper bound of the magnetic moment of 0.03~$\mu_{\text{B}}/\text{Ni}$ would be an underestimation for the $(1\,1\,L)$, $L$~=~odd position if they existed at 300~K due to a higher background as observed in Fig.~\ref{C_Rocking_11L} and the Debye-Waller factor due to the large temperature difference \cite{tazuke1997magnetic}. 

Several aspects of the magnetism in \LaNi{} are still unknown, but we will briefly speculate on the reason for for multiple magnetic transitions. First, we will start with a description of the magnetic structure at base temperature, and how AFM order can stabilize at relatively high temperatures with a small magnetic moment. At base temperature the magnetic structure is an uniaxial AFM in which all non-zero magnetic moments are along $\bm{c}$ with FM-aligned blocks stacked anti-parallel as shown in Fig.~\ref{Structure}. The structure predominantly has FM coupling between Ni except across the Ni5 position located at the position where the magnetic moment is zero. The magnetic moment is itinerant and likely forms from the sharp narrow peak in the DOS at the Fermi energy from $3d$ Ni contributions \cite{crivello2020relation,werwinski2018effect,singh2015electronic,werwinski2019effect}. In addition, the relatively large Rhodes-Wohlfarth ratio $\mu_{\text{c}}/\mu_{\text{sat}}=5.3$ \cite{ribeiro2022small} is consistent with an itinerant system. \LaNi{} sits close to a QCP in which both AFM and FM ground states are close in energy as demonstrated by both DFT results \cite{crivello2020relation} and experimental results that show with only 3\% Cu substitution \LaNi{} becomes FM. 

Phase B and C stabilize above the critical temperature for the low temperature phase A. The propagation vector of each of these phases is incommensurate, along $\bm{c}$, but with a larger periodicity. This means that the Ni magnetic moments are no longer ferromagnetic blocks. The incommensurate phases may be a result of frustration between FM and AFM interactions leading to an unrelated propagation vector which can result in either AFM or ferrimagnetic order. Hence, the observed in plane magnetic moment in phase B could be a consequence of competing exchange interactions and magnetic anisotropy of each non-symmetry related Ni site.

In order to have a more grounded understanding of the magnetism within \LaNi{} with its large unit cell much more experimental and theoretical work must be done. Further neutron scattering work as a function of field, and to fully determine the zero field structures would be very helpful. It is important to understand the electronic band structure of an itinerant magnet, so information on the propagation vector and the fully determined magnetic structures should be used to refine modeling for further DFT calculations. These could then be compared to recent ARPES results showing significant differences in the in-plane dispersion for the A, B, C, and PM phases\cite{lee2022electronic}. In addition, local probes such as NMR would greatly benefit efforts to understand the magnetic character of \LaNi{}.

\section{Conclusion}

We have identified three AFM phases in \LaNi{} in \LaNi{} at zero magnetic field using single crystal neutron diffraction with results summarized in Table~\ref{Tab:summary_phases}. These consisted of two incommensurate phases and one commensurate phases labeled the \ICc{}, \ICab{}, and \Cc{} with phase transitions at \TNICc{}~$=$~62.3(7), \TNICab{}~$=$~57.2(5), and \TNCc{}~$=$~42.8(5)~K, respectively. These three phase transitions consist of second-order phase transitions for the incommensurate phases and a first-order magnetic transition for the commensurate phase. The incommensurate phases have a propagation vector that increases with decreasing temperature, where the magnetic moment develops $\bm{\mu} \parallel \bm{c}$ for \TNICc{}~$>T>$~\TNCc{}, and $\bm{\mu} \perp \bm{c}$ for \TNICab{}~$>T>$~\TNCc{}. Below \TNCc{} the propagation vector becomes $(0\,0\,1)$, and the sequence of intensities for magnetic Bragg peaks is consistent the triangle-wave modulated spin-density wave shown in Fig~\ref{Structure}(c) predicted by recent DFT calculations \cite{crivello2020relation}.

\begin{acknowledgments}
	Work at Ames Laboratory was supported by the U.\,S.\ Department of Energy, Office of Basic Energy Sciences, Division of Materials Sciences \& Engineering. Ames Laboratory is operated for the U.\,S.\ Department of Energy by Iowa State under Contract No.\ DE-AC$02$-$07$CH$11358$. A portion of this research used resources at the High Flux Isotope Reactor, a U.\,S.\ Department of Energy, Office of Science User Facility operated by Oak Ridge National Laboratory.
\end{acknowledgments}

\bibliographystyle{apsrev4-2.bst}
\bibliography{La2Ni7AFM.bib}

\providecommand{\noopsort}[1]{}\providecommand{\singleletter}[1]{#1}%
\begin{thebibliography}{54}%
\makeatletter
\providecommand \@ifxundefined [1]{%
 \@ifx{#1\undefined}
}%
\providecommand \@ifnum [1]{%
 \ifnum #1\expandafter \@firstoftwo
 \else \expandafter \@secondoftwo
 \fi
}%
\providecommand \@ifx [1]{%
 \ifx #1\expandafter \@firstoftwo
 \else \expandafter \@secondoftwo
 \fi
}%
\providecommand \natexlab [1]{#1}%
\providecommand \enquote  [1]{``#1''}%
\providecommand \bibnamefont  [1]{#1}%
\providecommand \bibfnamefont [1]{#1}%
\providecommand \citenamefont [1]{#1}%
\providecommand \href@noop [0]{\@secondoftwo}%
\providecommand \href [0]{\begingroup \@sanitize@url \@href}%
\providecommand \@href[1]{\@@startlink{#1}\@@href}%
\providecommand \@@href[1]{\endgroup#1\@@endlink}%
\providecommand \@sanitize@url [0]{\catcode `\\12\catcode `\$12\catcode
  `\&12\catcode `\#12\catcode `\^12\catcode `\_12\catcode `\%12\relax}%
\providecommand \@@startlink[1]{}%
\providecommand \@@endlink[0]{}%
\providecommand \url  [0]{\begingroup\@sanitize@url \@url }%
\providecommand \@url [1]{\endgroup\@href {#1}{\urlprefix }}%
\providecommand \urlprefix  [0]{URL }%
\providecommand \Eprint [0]{\href }%
\providecommand \doibase [0]{https://doi.org/}%
\providecommand \selectlanguage [0]{\@gobble}%
\providecommand \bibinfo  [0]{\@secondoftwo}%
\providecommand \bibfield  [0]{\@secondoftwo}%
\providecommand \translation [1]{[#1]}%
\providecommand \BibitemOpen [0]{}%
\providecommand \bibitemStop [0]{}%
\providecommand \bibitemNoStop [0]{.\EOS\space}%
\providecommand \EOS [0]{\spacefactor3000\relax}%
\providecommand \BibitemShut  [1]{\csname bibitem#1\endcsname}%
\let\auto@bib@innerbib\@empty
\bibitem [{\citenamefont {Stewart}(2001)}]{stewart2001non}%
  \BibitemOpen
  \bibfield  {author} {\bibinfo {author} {\bibfnamefont {G.}~\bibnamefont
  {Stewart}},\ }\href@noop {} {\bibfield  {journal} {\bibinfo  {journal} {Rev.
  Mod. Phys.}\ }\textbf {\bibinfo {volume} {73}},\ \bibinfo {pages} {797}
  (\bibinfo {year} {2001})}\BibitemShut {NoStop}%
\bibitem [{\citenamefont {Canfield}\ and\ \citenamefont
  {Bud’ko}(2016)}]{canfield2016preserved}%
  \BibitemOpen
  \bibfield  {author} {\bibinfo {author} {\bibfnamefont {P.~C.}\ \bibnamefont
  {Canfield}}\ and\ \bibinfo {author} {\bibfnamefont {S.~L.}\ \bibnamefont
  {Bud’ko}},\ }\href@noop {} {\bibfield  {journal} {\bibinfo  {journal} {Rep.
  Prog. Phys.}\ }\textbf {\bibinfo {volume} {79}},\ \bibinfo {pages} {084506}
  (\bibinfo {year} {2016})}\BibitemShut {NoStop}%
\bibitem [{\citenamefont {Canfield}(2020)}]{canfield2020new}%
  \BibitemOpen
  \bibfield  {author} {\bibinfo {author} {\bibfnamefont {P.~C.}\ \bibnamefont
  {Canfield}},\ }\href@noop {} {\bibfield  {journal} {\bibinfo  {journal} {Rep.
  Prog. Phys.}\ }\textbf {\bibinfo {volume} {83}},\ \bibinfo {pages} {016501}
  (\bibinfo {year} {2020})}\BibitemShut {NoStop}%
\bibitem [{\citenamefont {Pfleiderer}\ \emph {et~al.}(2001)\citenamefont
  {Pfleiderer}, \citenamefont {Julian},\ and\ \citenamefont
  {Lonzarich}}]{pfleiderer2001non}%
  \BibitemOpen
  \bibfield  {author} {\bibinfo {author} {\bibfnamefont {C.}~\bibnamefont
  {Pfleiderer}}, \bibinfo {author} {\bibfnamefont {S.}~\bibnamefont {Julian}},\
  and\ \bibinfo {author} {\bibfnamefont {G.}~\bibnamefont {Lonzarich}},\
  }\href@noop {} {\bibfield  {journal} {\bibinfo  {journal} {Nature}\ }\textbf
  {\bibinfo {volume} {414}},\ \bibinfo {pages} {427} (\bibinfo {year}
  {2001})}\BibitemShut {NoStop}%
\bibitem [{\citenamefont {L{\'e}vy}\ \emph {et~al.}(2007)\citenamefont
  {L{\'e}vy}, \citenamefont {Sheikin},\ and\ \citenamefont
  {Huxley}}]{levy2007acute}%
  \BibitemOpen
  \bibfield  {author} {\bibinfo {author} {\bibfnamefont {F.}~\bibnamefont
  {L{\'e}vy}}, \bibinfo {author} {\bibfnamefont {I.}~\bibnamefont {Sheikin}},\
  and\ \bibinfo {author} {\bibfnamefont {A.}~\bibnamefont {Huxley}},\
  }\href@noop {} {\bibfield  {journal} {\bibinfo  {journal} {Nature Phys.}\
  }\textbf {\bibinfo {volume} {3}},\ \bibinfo {pages} {460} (\bibinfo {year}
  {2007})}\BibitemShut {NoStop}%
\bibitem [{\citenamefont {Uemura}\ \emph {et~al.}(2007)\citenamefont {Uemura},
  \citenamefont {Goko}, \citenamefont {Gat-Malureanu}, \citenamefont {Carlo},
  \citenamefont {Russo}, \citenamefont {Savici}, \citenamefont {Aczel},
  \citenamefont {MacDougall}, \citenamefont {Rodriguez}, \citenamefont {Luke}
  \emph {et~al.}}]{uemura2007phase}%
  \BibitemOpen
  \bibfield  {author} {\bibinfo {author} {\bibfnamefont {Y.}~\bibnamefont
  {Uemura}}, \bibinfo {author} {\bibfnamefont {T.}~\bibnamefont {Goko}},
  \bibinfo {author} {\bibfnamefont {I.}~\bibnamefont {Gat-Malureanu}}, \bibinfo
  {author} {\bibfnamefont {J.}~\bibnamefont {Carlo}}, \bibinfo {author}
  {\bibfnamefont {P.}~\bibnamefont {Russo}}, \bibinfo {author} {\bibfnamefont
  {A.}~\bibnamefont {Savici}}, \bibinfo {author} {\bibfnamefont
  {A.}~\bibnamefont {Aczel}}, \bibinfo {author} {\bibfnamefont
  {G.}~\bibnamefont {MacDougall}}, \bibinfo {author} {\bibfnamefont
  {J.}~\bibnamefont {Rodriguez}}, \bibinfo {author} {\bibfnamefont
  {G.}~\bibnamefont {Luke}}, \emph {et~al.},\ }\href@noop {} {\bibfield
  {journal} {\bibinfo  {journal} {Nature Phys.}\ }\textbf {\bibinfo {volume}
  {3}},\ \bibinfo {pages} {29} (\bibinfo {year} {2007})}\BibitemShut {NoStop}%
\bibitem [{\citenamefont {Huy}\ \emph {et~al.}(2007)\citenamefont {Huy},
  \citenamefont {Gasparini}, \citenamefont {De~Nijs}, \citenamefont {Huang},
  \citenamefont {Klaasse}, \citenamefont {Gortenmulder}, \citenamefont
  {de~Visser}, \citenamefont {Hamann}, \citenamefont {G{\"o}rlach},\ and\
  \citenamefont {L{\"o}hneysen}}]{huy2007superconductivity}%
  \BibitemOpen
  \bibfield  {author} {\bibinfo {author} {\bibfnamefont {N.}~\bibnamefont
  {Huy}}, \bibinfo {author} {\bibfnamefont {A.}~\bibnamefont {Gasparini}},
  \bibinfo {author} {\bibfnamefont {D.}~\bibnamefont {De~Nijs}}, \bibinfo
  {author} {\bibfnamefont {Y.}~\bibnamefont {Huang}}, \bibinfo {author}
  {\bibfnamefont {J.}~\bibnamefont {Klaasse}}, \bibinfo {author} {\bibfnamefont
  {T.}~\bibnamefont {Gortenmulder}}, \bibinfo {author} {\bibfnamefont
  {A.}~\bibnamefont {de~Visser}}, \bibinfo {author} {\bibfnamefont
  {A.}~\bibnamefont {Hamann}}, \bibinfo {author} {\bibfnamefont
  {T.}~\bibnamefont {G{\"o}rlach}},\ and\ \bibinfo {author} {\bibfnamefont
  {H.~v.}\ \bibnamefont {L{\"o}hneysen}},\ }\href@noop {} {\bibfield  {journal}
  {\bibinfo  {journal} {Phys. Rev. Lett.}\ }\textbf {\bibinfo {volume} {99}},\
  \bibinfo {pages} {067006} (\bibinfo {year} {2007})}\BibitemShut {NoStop}%
\bibitem [{\citenamefont {Cheng}\ \emph {et~al.}(2015)\citenamefont {Cheng},
  \citenamefont {Matsubayashi}, \citenamefont {Wu}, \citenamefont {Sun},
  \citenamefont {Lin}, \citenamefont {Luo},\ and\ \citenamefont
  {Uwatoko}}]{cheng2015pressure}%
  \BibitemOpen
  \bibfield  {author} {\bibinfo {author} {\bibfnamefont {J.-G.}\ \bibnamefont
  {Cheng}}, \bibinfo {author} {\bibfnamefont {K.}~\bibnamefont {Matsubayashi}},
  \bibinfo {author} {\bibfnamefont {W.}~\bibnamefont {Wu}}, \bibinfo {author}
  {\bibfnamefont {J.-P.}\ \bibnamefont {Sun}}, \bibinfo {author} {\bibfnamefont
  {F.-K.}\ \bibnamefont {Lin}}, \bibinfo {author} {\bibfnamefont {J.-L.}\
  \bibnamefont {Luo}},\ and\ \bibinfo {author} {\bibfnamefont {Y.}~\bibnamefont
  {Uwatoko}},\ }\href@noop {} {\bibfield  {journal} {\bibinfo  {journal} {Phys.
  Rev. Lett.}\ }\textbf {\bibinfo {volume} {114}},\ \bibinfo {pages} {117001}
  (\bibinfo {year} {2015})}\BibitemShut {NoStop}%
\bibitem [{\citenamefont {Brando}\ \emph {et~al.}(2016)\citenamefont {Brando},
  \citenamefont {Belitz}, \citenamefont {Grosche},\ and\ \citenamefont
  {Kirkpatrick}}]{brando2016metallic}%
  \BibitemOpen
  \bibfield  {author} {\bibinfo {author} {\bibfnamefont {M.}~\bibnamefont
  {Brando}}, \bibinfo {author} {\bibfnamefont {D.}~\bibnamefont {Belitz}},
  \bibinfo {author} {\bibfnamefont {F.~M.}\ \bibnamefont {Grosche}},\ and\
  \bibinfo {author} {\bibfnamefont {T.~R.}\ \bibnamefont {Kirkpatrick}},\
  }\href@noop {} {\bibfield  {journal} {\bibinfo  {journal} {Rev. Mod. Phys.}\
  }\textbf {\bibinfo {volume} {88}},\ \bibinfo {pages} {025006} (\bibinfo
  {year} {2016})}\BibitemShut {NoStop}%
\bibitem [{\citenamefont {Ran}\ \emph {et~al.}(2019)\citenamefont {Ran},
  \citenamefont {Eckberg}, \citenamefont {Ding}, \citenamefont {Furukawa},
  \citenamefont {Metz}, \citenamefont {Saha}, \citenamefont {Liu},
  \citenamefont {Zic}, \citenamefont {Kim}, \citenamefont {Paglione} \emph
  {et~al.}}]{ran2019nearly}%
  \BibitemOpen
  \bibfield  {author} {\bibinfo {author} {\bibfnamefont {S.}~\bibnamefont
  {Ran}}, \bibinfo {author} {\bibfnamefont {C.}~\bibnamefont {Eckberg}},
  \bibinfo {author} {\bibfnamefont {Q.-P.}\ \bibnamefont {Ding}}, \bibinfo
  {author} {\bibfnamefont {Y.}~\bibnamefont {Furukawa}}, \bibinfo {author}
  {\bibfnamefont {T.}~\bibnamefont {Metz}}, \bibinfo {author} {\bibfnamefont
  {S.~R.}\ \bibnamefont {Saha}}, \bibinfo {author} {\bibfnamefont {I.-L.}\
  \bibnamefont {Liu}}, \bibinfo {author} {\bibfnamefont {M.}~\bibnamefont
  {Zic}}, \bibinfo {author} {\bibfnamefont {H.}~\bibnamefont {Kim}}, \bibinfo
  {author} {\bibfnamefont {J.}~\bibnamefont {Paglione}}, \emph {et~al.},\
  }\href@noop {} {\bibfield  {journal} {\bibinfo  {journal} {Science}\ }\textbf
  {\bibinfo {volume} {365}},\ \bibinfo {pages} {684} (\bibinfo {year}
  {2019})}\BibitemShut {NoStop}%
\bibitem [{\citenamefont {Belitz}\ \emph {et~al.}(1997)\citenamefont {Belitz},
  \citenamefont {Kirkpatrick},\ and\ \citenamefont
  {Vojta}}]{belitz1997nonanalytic}%
  \BibitemOpen
  \bibfield  {author} {\bibinfo {author} {\bibfnamefont {D.}~\bibnamefont
  {Belitz}}, \bibinfo {author} {\bibfnamefont {T.~R.}\ \bibnamefont
  {Kirkpatrick}},\ and\ \bibinfo {author} {\bibfnamefont {T.}~\bibnamefont
  {Vojta}},\ }\href@noop {} {\bibfield  {journal} {\bibinfo  {journal} {Phys.
  Rev. B}\ }\textbf {\bibinfo {volume} {55}},\ \bibinfo {pages} {9452}
  (\bibinfo {year} {1997})}\BibitemShut {NoStop}%
\bibitem [{\citenamefont {Shibauchi}\ \emph {et~al.}(2014)\citenamefont
  {Shibauchi}, \citenamefont {Carrington},\ and\ \citenamefont
  {Matsuda}}]{shibauchi2014quantum}%
  \BibitemOpen
  \bibfield  {author} {\bibinfo {author} {\bibfnamefont {T.}~\bibnamefont
  {Shibauchi}}, \bibinfo {author} {\bibfnamefont {A.}~\bibnamefont
  {Carrington}},\ and\ \bibinfo {author} {\bibfnamefont {Y.}~\bibnamefont
  {Matsuda}},\ }\href@noop {} {\bibfield  {journal} {\bibinfo  {journal} {Annu.
  Rev. Condens. Matt. Phys.}\ }\textbf {\bibinfo {volume} {5}},\ \bibinfo
  {pages} {113} (\bibinfo {year} {2014})}\BibitemShut {NoStop}%
\bibitem [{\citenamefont {Gegenwart}\ \emph {et~al.}(2008)\citenamefont
  {Gegenwart}, \citenamefont {Si},\ and\ \citenamefont
  {Steglich}}]{gegenwart2008quantum}%
  \BibitemOpen
  \bibfield  {author} {\bibinfo {author} {\bibfnamefont {P.}~\bibnamefont
  {Gegenwart}}, \bibinfo {author} {\bibfnamefont {Q.}~\bibnamefont {Si}},\ and\
  \bibinfo {author} {\bibfnamefont {F.}~\bibnamefont {Steglich}},\ }\href@noop
  {} {\bibfield  {journal} {\bibinfo  {journal} {Nature Phys.}\ }\textbf
  {\bibinfo {volume} {4}},\ \bibinfo {pages} {186} (\bibinfo {year}
  {2008})}\BibitemShut {NoStop}%
\bibitem [{\citenamefont {Moriya}\ \emph {et~al.}(1992)\citenamefont {Moriya},
  \citenamefont {Takahashi},\ and\ \citenamefont
  {Ueda}}]{moriya1992antiferromagnetic}%
  \BibitemOpen
  \bibfield  {author} {\bibinfo {author} {\bibfnamefont {T.}~\bibnamefont
  {Moriya}}, \bibinfo {author} {\bibfnamefont {Y.}~\bibnamefont {Takahashi}},\
  and\ \bibinfo {author} {\bibfnamefont {K.}~\bibnamefont {Ueda}},\ }\href@noop
  {} {\bibfield  {journal} {\bibinfo  {journal} {J. Magn. Magn. Mater}\
  }\textbf {\bibinfo {volume} {104}},\ \bibinfo {pages} {456} (\bibinfo {year}
  {1992})}\BibitemShut {NoStop}%
\bibitem [{\citenamefont {Ishigaki}\ and\ \citenamefont
  {Moriya}(1998)}]{ishigaki1998theory}%
  \BibitemOpen
  \bibfield  {author} {\bibinfo {author} {\bibfnamefont {A.}~\bibnamefont
  {Ishigaki}}\ and\ \bibinfo {author} {\bibfnamefont {T.}~\bibnamefont
  {Moriya}},\ }\href@noop {} {\bibfield  {journal} {\bibinfo  {journal} {J.
  Phys. Soc. Japan}\ }\textbf {\bibinfo {volume} {67}},\ \bibinfo {pages}
  {3924} (\bibinfo {year} {1998})}\BibitemShut {NoStop}%
\bibitem [{\citenamefont {Moriya}\ and\ \citenamefont
  {Ueda}(2003)}]{moriya2003antiferromagnetic}%
  \BibitemOpen
  \bibfield  {author} {\bibinfo {author} {\bibfnamefont {T.}~\bibnamefont
  {Moriya}}\ and\ \bibinfo {author} {\bibfnamefont {K.}~\bibnamefont {Ueda}},\
  }\href@noop {} {\bibfield  {journal} {\bibinfo  {journal} {Reports on
  Progress in Physics}\ }\textbf {\bibinfo {volume} {66}},\ \bibinfo {pages}
  {1299} (\bibinfo {year} {2003})}\BibitemShut {NoStop}%
\bibitem [{\citenamefont {Taufour}\ \emph {et~al.}(2016)\citenamefont
  {Taufour}, \citenamefont {Kaluarachchi}, \citenamefont {Khasanov},
  \citenamefont {Nguyen}, \citenamefont {Guguchia}, \citenamefont {Biswas},
  \citenamefont {Bonf{\`a}}, \citenamefont {De~Renzi}, \citenamefont {Lin},
  \citenamefont {Kim}, \citenamefont {Mun}, \citenamefont {Kim}, \citenamefont
  {Furukawa}, \citenamefont {Wang}, \citenamefont {Ho}, \citenamefont {Budko},\
  and\ \citenamefont {Canfield}}]{taufour2016ferromagnetic}%
  \BibitemOpen
  \bibfield  {author} {\bibinfo {author} {\bibfnamefont {V.}~\bibnamefont
  {Taufour}}, \bibinfo {author} {\bibfnamefont {U.~S.}\ \bibnamefont
  {Kaluarachchi}}, \bibinfo {author} {\bibfnamefont {R.}~\bibnamefont
  {Khasanov}}, \bibinfo {author} {\bibfnamefont {M.~C.}\ \bibnamefont
  {Nguyen}}, \bibinfo {author} {\bibfnamefont {Z.}~\bibnamefont {Guguchia}},
  \bibinfo {author} {\bibfnamefont {P.~K.}\ \bibnamefont {Biswas}}, \bibinfo
  {author} {\bibfnamefont {P.}~\bibnamefont {Bonf{\`a}}}, \bibinfo {author}
  {\bibfnamefont {R.}~\bibnamefont {De~Renzi}}, \bibinfo {author}
  {\bibfnamefont {X.}~\bibnamefont {Lin}}, \bibinfo {author} {\bibfnamefont
  {S.~K.}\ \bibnamefont {Kim}}, \bibinfo {author} {\bibfnamefont {E.~D.}\
  \bibnamefont {Mun}}, \bibinfo {author} {\bibfnamefont {H.}~\bibnamefont
  {Kim}}, \bibinfo {author} {\bibfnamefont {Y.}~\bibnamefont {Furukawa}},
  \bibinfo {author} {\bibfnamefont {C.~Z.}\ \bibnamefont {Wang}}, \bibinfo
  {author} {\bibfnamefont {K.~M.}\ \bibnamefont {Ho}}, \bibinfo {author}
  {\bibfnamefont {S.~L.}\ \bibnamefont {Budko}},\ and\ \bibinfo {author}
  {\bibfnamefont {P.~C.}\ \bibnamefont {Canfield}},\ }\href@noop {} {\bibfield
  {journal} {\bibinfo  {journal} {Phys. Rev. Lett.}\ }\textbf {\bibinfo
  {volume} {117}},\ \bibinfo {pages} {037207} (\bibinfo {year}
  {2016})}\BibitemShut {NoStop}%
\bibitem [{\citenamefont {Kaluarachchi}\ \emph {et~al.}(2017)\citenamefont
  {Kaluarachchi}, \citenamefont {Bud’ko}, \citenamefont {Canfield},\ and\
  \citenamefont {Taufour}}]{kaluarachchi2017tricritical}%
  \BibitemOpen
  \bibfield  {author} {\bibinfo {author} {\bibfnamefont {U.~S.}\ \bibnamefont
  {Kaluarachchi}}, \bibinfo {author} {\bibfnamefont {S.~L.}\ \bibnamefont
  {Bud’ko}}, \bibinfo {author} {\bibfnamefont {P.~C.}\ \bibnamefont
  {Canfield}},\ and\ \bibinfo {author} {\bibfnamefont {V.}~\bibnamefont
  {Taufour}},\ }\href@noop {} {\bibfield  {journal} {\bibinfo  {journal}
  {Nature Comm.}\ }\textbf {\bibinfo {volume} {8}},\ \bibinfo {pages} {1}
  (\bibinfo {year} {2017})}\BibitemShut {NoStop}%
\bibitem [{\citenamefont {Gati}\ \emph {et~al.}(2021)\citenamefont {Gati},
  \citenamefont {Wilde}, \citenamefont {Khasanov}, \citenamefont {Xiang},
  \citenamefont {Dissanayake}, \citenamefont {Gupta}, \citenamefont {Matsuda},
  \citenamefont {Ye}, \citenamefont {Haberl}, \citenamefont {Kaluarachchi},
  \citenamefont {McQueeney}, \citenamefont {Kreyssig}, \citenamefont {Budko},\
  and\ \citenamefont {Canfield}}]{gati2021formation}%
  \BibitemOpen
  \bibfield  {author} {\bibinfo {author} {\bibfnamefont {E.}~\bibnamefont
  {Gati}}, \bibinfo {author} {\bibfnamefont {J.~M.}\ \bibnamefont {Wilde}},
  \bibinfo {author} {\bibfnamefont {R.}~\bibnamefont {Khasanov}}, \bibinfo
  {author} {\bibfnamefont {L.}~\bibnamefont {Xiang}}, \bibinfo {author}
  {\bibfnamefont {S.}~\bibnamefont {Dissanayake}}, \bibinfo {author}
  {\bibfnamefont {R.}~\bibnamefont {Gupta}}, \bibinfo {author} {\bibfnamefont
  {M.}~\bibnamefont {Matsuda}}, \bibinfo {author} {\bibfnamefont
  {F.}~\bibnamefont {Ye}}, \bibinfo {author} {\bibfnamefont {B.}~\bibnamefont
  {Haberl}}, \bibinfo {author} {\bibfnamefont {U.}~\bibnamefont
  {Kaluarachchi}}, \bibinfo {author} {\bibfnamefont {R.~J.}\ \bibnamefont
  {McQueeney}}, \bibinfo {author} {\bibfnamefont {A.}~\bibnamefont {Kreyssig}},
  \bibinfo {author} {\bibfnamefont {S.~L.}\ \bibnamefont {Budko}},\ and\
  \bibinfo {author} {\bibfnamefont {P.~C.}\ \bibnamefont {Canfield}},\
  }\href@noop {} {\bibfield  {journal} {\bibinfo  {journal} {Phys. Rev. B}\
  }\textbf {\bibinfo {volume} {103}},\ \bibinfo {pages} {075111} (\bibinfo
  {year} {2021})}\BibitemShut {NoStop}%
\bibitem [{\citenamefont {Xiang}\ \emph {et~al.}(2021)\citenamefont {Xiang},
  \citenamefont {Gati}, \citenamefont {Bud'ko}, \citenamefont {Saunders},\ and\
  \citenamefont {Canfield}}]{xiang2021avoided}%
  \BibitemOpen
  \bibfield  {author} {\bibinfo {author} {\bibfnamefont {L.}~\bibnamefont
  {Xiang}}, \bibinfo {author} {\bibfnamefont {E.}~\bibnamefont {Gati}},
  \bibinfo {author} {\bibfnamefont {S.~L.}\ \bibnamefont {Bud'ko}}, \bibinfo
  {author} {\bibfnamefont {S.~M.}\ \bibnamefont {Saunders}},\ and\ \bibinfo
  {author} {\bibfnamefont {P.~C.}\ \bibnamefont {Canfield}},\ }\href@noop {}
  {\bibfield  {journal} {\bibinfo  {journal} {Phys. Rev. B}\ }\textbf {\bibinfo
  {volume} {103}},\ \bibinfo {pages} {054419} (\bibinfo {year}
  {2021})}\BibitemShut {NoStop}%
\bibitem [{\citenamefont {Matthias}\ \emph {et~al.}(1978)\citenamefont
  {Matthias}, \citenamefont {Giorgi}, \citenamefont {Struebing},\ and\
  \citenamefont {Smith}}]{matthias1978itinerant}%
  \BibitemOpen
  \bibfield  {author} {\bibinfo {author} {\bibfnamefont {B.}~\bibnamefont
  {Matthias}}, \bibinfo {author} {\bibfnamefont {A.}~\bibnamefont {Giorgi}},
  \bibinfo {author} {\bibfnamefont {V.}~\bibnamefont {Struebing}},\ and\
  \bibinfo {author} {\bibfnamefont {J.}~\bibnamefont {Smith}},\ }\href@noop {}
  {\bibfield  {journal} {\bibinfo  {journal} {Journal de Physique Lettres}\
  }\textbf {\bibinfo {volume} {39}},\ \bibinfo {pages} {441} (\bibinfo {year}
  {1978})}\BibitemShut {NoStop}%
\bibitem [{\citenamefont {Povzner}\ and\ \citenamefont
  {Likhachev}(1995)}]{povzner1995spin}%
  \BibitemOpen
  \bibfield  {author} {\bibinfo {author} {\bibfnamefont {A.}~\bibnamefont
  {Povzner}}\ and\ \bibinfo {author} {\bibfnamefont {D.}~\bibnamefont
  {Likhachev}},\ }\href@noop {} {\bibfield  {journal} {\bibinfo  {journal}
  {International Journal of Modern Physics B}\ }\textbf {\bibinfo {volume}
  {9}},\ \bibinfo {pages} {1171} (\bibinfo {year} {1995})}\BibitemShut
  {NoStop}%
\bibitem [{\citenamefont {Torun}\ \emph {et~al.}(2016)\citenamefont {Torun},
  \citenamefont {Janner},\ and\ \citenamefont {De~Groot}}]{torun2016origin}%
  \BibitemOpen
  \bibfield  {author} {\bibinfo {author} {\bibfnamefont {E.}~\bibnamefont
  {Torun}}, \bibinfo {author} {\bibfnamefont {A.}~\bibnamefont {Janner}},\ and\
  \bibinfo {author} {\bibfnamefont {R.}~\bibnamefont {De~Groot}},\ }\href@noop
  {} {\bibfield  {journal} {\bibinfo  {journal} {Journal of Physics: Condensed
  Matter}\ }\textbf {\bibinfo {volume} {28}},\ \bibinfo {pages} {065501}
  (\bibinfo {year} {2016})}\BibitemShut {NoStop}%
\bibitem [{\citenamefont {Svanidze}\ \emph {et~al.}(2015)\citenamefont
  {Svanidze}, \citenamefont {Wang}, \citenamefont {Besara}, \citenamefont
  {Liu}, \citenamefont {Huang}, \citenamefont {Siegrist}, \citenamefont
  {Frandsen}, \citenamefont {Lynn}, \citenamefont {Nevidomskyy}, \citenamefont
  {Gam{\.z}a} \emph {et~al.}}]{svanidze2015itinerant}%
  \BibitemOpen
  \bibfield  {author} {\bibinfo {author} {\bibfnamefont {E.}~\bibnamefont
  {Svanidze}}, \bibinfo {author} {\bibfnamefont {J.~K.}\ \bibnamefont {Wang}},
  \bibinfo {author} {\bibfnamefont {T.}~\bibnamefont {Besara}}, \bibinfo
  {author} {\bibfnamefont {L.}~\bibnamefont {Liu}}, \bibinfo {author}
  {\bibfnamefont {Q.}~\bibnamefont {Huang}}, \bibinfo {author} {\bibfnamefont
  {T.}~\bibnamefont {Siegrist}}, \bibinfo {author} {\bibfnamefont
  {B.}~\bibnamefont {Frandsen}}, \bibinfo {author} {\bibfnamefont {J.~W.}\
  \bibnamefont {Lynn}}, \bibinfo {author} {\bibfnamefont {A.~H.}\ \bibnamefont
  {Nevidomskyy}}, \bibinfo {author} {\bibfnamefont {M.~B.}\ \bibnamefont
  {Gam{\.z}a}}, \emph {et~al.},\ }\href@noop {} {\bibfield  {journal} {\bibinfo
   {journal} {Nature Comm.}\ }\textbf {\bibinfo {volume} {6}},\ \bibinfo
  {pages} {1} (\bibinfo {year} {2015})}\BibitemShut {NoStop}%
\bibitem [{\citenamefont {Goh}\ and\ \citenamefont
  {Pickett}(2016)}]{goh2016mechanism}%
  \BibitemOpen
  \bibfield  {author} {\bibinfo {author} {\bibfnamefont {W.~F.}\ \bibnamefont
  {Goh}}\ and\ \bibinfo {author} {\bibfnamefont {W.~E.}\ \bibnamefont
  {Pickett}},\ }\href@noop {} {\bibfield  {journal} {\bibinfo  {journal} {EPL
  (Europhysics Letters)}\ }\textbf {\bibinfo {volume} {116}},\ \bibinfo {pages}
  {27004} (\bibinfo {year} {2016})}\BibitemShut {NoStop}%
\bibitem [{\citenamefont {Goh}\ and\ \citenamefont
  {Pickett}(2017)}]{goh2017competing}%
  \BibitemOpen
  \bibfield  {author} {\bibinfo {author} {\bibfnamefont {W.~F.}\ \bibnamefont
  {Goh}}\ and\ \bibinfo {author} {\bibfnamefont {W.~E.}\ \bibnamefont
  {Pickett}},\ }\href@noop {} {\bibfield  {journal} {\bibinfo  {journal} {Phys.
  Rev. B}\ }\textbf {\bibinfo {volume} {95}},\ \bibinfo {pages} {205124}
  (\bibinfo {year} {2017})}\BibitemShut {NoStop}%
\bibitem [{\citenamefont {Mathew}\ \emph {et~al.}(2019)\citenamefont {Mathew},
  \citenamefont {Goh},\ and\ \citenamefont {Pickett}}]{mathew2019probing}%
  \BibitemOpen
  \bibfield  {author} {\bibinfo {author} {\bibfnamefont {M.}~\bibnamefont
  {Mathew}}, \bibinfo {author} {\bibfnamefont {W.~F.}\ \bibnamefont {Goh}},\
  and\ \bibinfo {author} {\bibfnamefont {W.~E.}\ \bibnamefont {Pickett}},\
  }\href@noop {} {\bibfield  {journal} {\bibinfo  {journal} {Journal of
  Physics: Condensed Matter}\ }\textbf {\bibinfo {volume} {31}},\ \bibinfo
  {pages} {074005} (\bibinfo {year} {2019})}\BibitemShut {NoStop}%
\bibitem [{\citenamefont {Sangeetha}\ \emph {et~al.}(2019)\citenamefont
  {Sangeetha}, \citenamefont {Wang}, \citenamefont {Smirnov}, \citenamefont
  {Smetana}, \citenamefont {Mudring}, \citenamefont {Johnson}, \citenamefont
  {Tanatar}, \citenamefont {Prozorov},\ and\ \citenamefont
  {Johnston}}]{sangeetha2019non}%
  \BibitemOpen
  \bibfield  {author} {\bibinfo {author} {\bibfnamefont {N.~S.}\ \bibnamefont
  {Sangeetha}}, \bibinfo {author} {\bibfnamefont {L.-L.}\ \bibnamefont {Wang}},
  \bibinfo {author} {\bibfnamefont {A.~V.}\ \bibnamefont {Smirnov}}, \bibinfo
  {author} {\bibfnamefont {V.}~\bibnamefont {Smetana}}, \bibinfo {author}
  {\bibfnamefont {A.-V.}\ \bibnamefont {Mudring}}, \bibinfo {author}
  {\bibfnamefont {D.~D.}\ \bibnamefont {Johnson}}, \bibinfo {author}
  {\bibfnamefont {M.~A.}\ \bibnamefont {Tanatar}}, \bibinfo {author}
  {\bibfnamefont {R.}~\bibnamefont {Prozorov}},\ and\ \bibinfo {author}
  {\bibfnamefont {D.~C.}\ \bibnamefont {Johnston}},\ }\href@noop {} {\bibfield
  {journal} {\bibinfo  {journal} {Phys. Rev. B}\ }\textbf {\bibinfo {volume}
  {100}},\ \bibinfo {pages} {094447} (\bibinfo {year} {2019})}\BibitemShut
  {NoStop}%
\bibitem [{\citenamefont {Wilde}\ \emph {et~al.}(2019)\citenamefont {Wilde},
  \citenamefont {Kreyssig}, \citenamefont {Vaknin}, \citenamefont {Sangeetha},
  \citenamefont {Li}, \citenamefont {Tian}, \citenamefont {Orth}, \citenamefont
  {Johnston}, \citenamefont {Ueland},\ and\ \citenamefont
  {McQueeney}}]{wilde2019helical}%
  \BibitemOpen
  \bibfield  {author} {\bibinfo {author} {\bibfnamefont {J.~M.}\ \bibnamefont
  {Wilde}}, \bibinfo {author} {\bibfnamefont {A.}~\bibnamefont {Kreyssig}},
  \bibinfo {author} {\bibfnamefont {D.}~\bibnamefont {Vaknin}}, \bibinfo
  {author} {\bibfnamefont {N.~S.}\ \bibnamefont {Sangeetha}}, \bibinfo {author}
  {\bibfnamefont {B.}~\bibnamefont {Li}}, \bibinfo {author} {\bibfnamefont
  {W.}~\bibnamefont {Tian}}, \bibinfo {author} {\bibfnamefont {P.~P.}\
  \bibnamefont {Orth}}, \bibinfo {author} {\bibfnamefont {D.~C.}\ \bibnamefont
  {Johnston}}, \bibinfo {author} {\bibfnamefont {B.~G.}\ \bibnamefont
  {Ueland}},\ and\ \bibinfo {author} {\bibfnamefont {R.~J.}\ \bibnamefont
  {McQueeney}},\ }\href@noop {} {\bibfield  {journal} {\bibinfo  {journal}
  {Phys. Rev. B}\ }\textbf {\bibinfo {volume} {100}},\ \bibinfo {pages}
  {161113(R)} (\bibinfo {year} {2019})}\BibitemShut {NoStop}%
\bibitem [{\citenamefont {Li}\ \emph {et~al.}(2019{\natexlab{a}})\citenamefont
  {Li}, \citenamefont {Liu}, \citenamefont {Xu}, \citenamefont {Song},
  \citenamefont {Huang}, \citenamefont {Shen}, \citenamefont {Ma},
  \citenamefont {Li}, \citenamefont {Chi}, \citenamefont {Frontzek},
  \citenamefont {Cao}, \citenamefont {Huang}, \citenamefont {Wang},
  \citenamefont {Xie}, \citenamefont {Zhang}, \citenamefont {Rong},
  \citenamefont {Shelton}, \citenamefont {Young}, \citenamefont {DiTusa},\ and\
  \citenamefont {Dai}}]{li2019flat}%
  \BibitemOpen
  \bibfield  {author} {\bibinfo {author} {\bibfnamefont {Y.}~\bibnamefont
  {Li}}, \bibinfo {author} {\bibfnamefont {Z.}~\bibnamefont {Liu}}, \bibinfo
  {author} {\bibfnamefont {Z.}~\bibnamefont {Xu}}, \bibinfo {author}
  {\bibfnamefont {Y.}~\bibnamefont {Song}}, \bibinfo {author} {\bibfnamefont
  {Y.}~\bibnamefont {Huang}}, \bibinfo {author} {\bibfnamefont
  {D.}~\bibnamefont {Shen}}, \bibinfo {author} {\bibfnamefont {N.}~\bibnamefont
  {Ma}}, \bibinfo {author} {\bibfnamefont {A.}~\bibnamefont {Li}}, \bibinfo
  {author} {\bibfnamefont {S.}~\bibnamefont {Chi}}, \bibinfo {author}
  {\bibfnamefont {M.}~\bibnamefont {Frontzek}}, \bibinfo {author}
  {\bibfnamefont {H.}~\bibnamefont {Cao}}, \bibinfo {author} {\bibfnamefont
  {Q.}~\bibnamefont {Huang}}, \bibinfo {author} {\bibfnamefont
  {W.}~\bibnamefont {Wang}}, \bibinfo {author} {\bibfnamefont {Y.}~\bibnamefont
  {Xie}}, \bibinfo {author} {\bibfnamefont {R.}~\bibnamefont {Zhang}}, \bibinfo
  {author} {\bibfnamefont {Y.}~\bibnamefont {Rong}}, \bibinfo {author}
  {\bibfnamefont {W.~A.}\ \bibnamefont {Shelton}}, \bibinfo {author}
  {\bibfnamefont {D.~P.}\ \bibnamefont {Young}}, \bibinfo {author}
  {\bibfnamefont {J.~F.}\ \bibnamefont {DiTusa}},\ and\ \bibinfo {author}
  {\bibfnamefont {P.}~\bibnamefont {Dai}},\ }\href@noop {} {\bibfield
  {journal} {\bibinfo  {journal} {Phys. Rev. B}\ }\textbf {\bibinfo {volume}
  {100}},\ \bibinfo {pages} {094446} (\bibinfo {year}
  {2019}{\natexlab{a}})}\BibitemShut {NoStop}%
\bibitem [{\citenamefont {Li}\ \emph {et~al.}(2019{\natexlab{b}})\citenamefont
  {Li}, \citenamefont {Sizyuk}, \citenamefont {Sangeetha}, \citenamefont
  {Wilde}, \citenamefont {Das}, \citenamefont {Tian}, \citenamefont {Johnston},
  \citenamefont {Goldman}, \citenamefont {Kreyssig}, \citenamefont {Orth},
  \citenamefont {McQueeney},\ and\ \citenamefont
  {Ueland}}]{li2019antiferromagnetic}%
  \BibitemOpen
  \bibfield  {author} {\bibinfo {author} {\bibfnamefont {B.}~\bibnamefont
  {Li}}, \bibinfo {author} {\bibfnamefont {Y.}~\bibnamefont {Sizyuk}}, \bibinfo
  {author} {\bibfnamefont {N.~S.}\ \bibnamefont {Sangeetha}}, \bibinfo {author}
  {\bibfnamefont {J.~M.}\ \bibnamefont {Wilde}}, \bibinfo {author}
  {\bibfnamefont {P.}~\bibnamefont {Das}}, \bibinfo {author} {\bibfnamefont
  {W.}~\bibnamefont {Tian}}, \bibinfo {author} {\bibfnamefont {D.~C.}\
  \bibnamefont {Johnston}}, \bibinfo {author} {\bibfnamefont {A.~I.}\
  \bibnamefont {Goldman}}, \bibinfo {author} {\bibfnamefont {A.}~\bibnamefont
  {Kreyssig}}, \bibinfo {author} {\bibfnamefont {P.~P.}\ \bibnamefont {Orth}},
  \bibinfo {author} {\bibfnamefont {R.~J.}\ \bibnamefont {McQueeney}},\ and\
  \bibinfo {author} {\bibfnamefont {B.~G.}\ \bibnamefont {Ueland}},\
  }\href@noop {} {\bibfield  {journal} {\bibinfo  {journal} {Phys. Rev. B}\
  }\textbf {\bibinfo {volume} {100}},\ \bibinfo {pages} {024415} (\bibinfo
  {year} {2019}{\natexlab{b}})}\BibitemShut {NoStop}%
\bibitem [{\citenamefont {Kadowaki}\ \emph {et~al.}(1982)\citenamefont
  {Kadowaki}, \citenamefont {Okuda},\ and\ \citenamefont
  {Date}}]{kadowaki1982magnetization}%
  \BibitemOpen
  \bibfield  {author} {\bibinfo {author} {\bibfnamefont {K.}~\bibnamefont
  {Kadowaki}}, \bibinfo {author} {\bibfnamefont {K.}~\bibnamefont {Okuda}},\
  and\ \bibinfo {author} {\bibfnamefont {M.}~\bibnamefont {Date}},\ }\href@noop
  {} {\bibfield  {journal} {\bibinfo  {journal} {J. Phys. Soc. Japan}\ }\textbf
  {\bibinfo {volume} {51}},\ \bibinfo {pages} {2433} (\bibinfo {year}
  {1982})}\BibitemShut {NoStop}%
\bibitem [{\citenamefont {Stishov}\ and\ \citenamefont
  {Petrova}(2011)}]{stishov2011itinerant}%
  \BibitemOpen
  \bibfield  {author} {\bibinfo {author} {\bibfnamefont {S.~M.}\ \bibnamefont
  {Stishov}}\ and\ \bibinfo {author} {\bibfnamefont {A.~E.}\ \bibnamefont
  {Petrova}},\ }\href@noop {} {\bibfield  {journal} {\bibinfo  {journal}
  {Physics-Uspekhi}\ }\textbf {\bibinfo {volume} {54}},\ \bibinfo {pages}
  {1117} (\bibinfo {year} {2011})}\BibitemShut {NoStop}%
\bibitem [{\citenamefont {Avila}\ \emph {et~al.}(2004)\citenamefont {Avila},
  \citenamefont {Bud'ko},\ and\ \citenamefont
  {Canfield}}]{avila2004anisotropic}%
  \BibitemOpen
  \bibfield  {author} {\bibinfo {author} {\bibfnamefont {M.~A.}\ \bibnamefont
  {Avila}}, \bibinfo {author} {\bibfnamefont {S.~L.}\ \bibnamefont {Bud'ko}},\
  and\ \bibinfo {author} {\bibfnamefont {P.~C.}\ \bibnamefont {Canfield}},\
  }\href@noop {} {\bibfield  {journal} {\bibinfo  {journal} {J. Magn. Magn.
  Mater.}\ }\textbf {\bibinfo {volume} {270}},\ \bibinfo {pages} {51} (\bibinfo
  {year} {2004})}\BibitemShut {NoStop}%
\bibitem [{\citenamefont {Fujiwara}\ \emph {et~al.}(2007)\citenamefont
  {Fujiwara}, \citenamefont {Aso}, \citenamefont {Yamamoto}, \citenamefont
  {Hedo}, \citenamefont {Saiga}, \citenamefont {Nishi}, \citenamefont
  {Uwatoko},\ and\ \citenamefont {Hirota}}]{fujiwara2007pressure}%
  \BibitemOpen
  \bibfield  {author} {\bibinfo {author} {\bibfnamefont {T.}~\bibnamefont
  {Fujiwara}}, \bibinfo {author} {\bibfnamefont {N.}~\bibnamefont {Aso}},
  \bibinfo {author} {\bibfnamefont {H.}~\bibnamefont {Yamamoto}}, \bibinfo
  {author} {\bibfnamefont {M.}~\bibnamefont {Hedo}}, \bibinfo {author}
  {\bibfnamefont {Y.}~\bibnamefont {Saiga}}, \bibinfo {author} {\bibfnamefont
  {M.}~\bibnamefont {Nishi}}, \bibinfo {author} {\bibfnamefont
  {Y.}~\bibnamefont {Uwatoko}},\ and\ \bibinfo {author} {\bibfnamefont
  {K.}~\bibnamefont {Hirota}},\ }\href@noop {} {\bibfield  {journal} {\bibinfo
  {journal} {J. Phys. Soc. Japan}\ }\textbf {\bibinfo {volume} {76}},\ \bibinfo
  {pages} {60} (\bibinfo {year} {2007})}\BibitemShut {NoStop}%
\bibitem [{\citenamefont {Buschow}(1983)}]{buschow1983magnetic}%
  \BibitemOpen
  \bibfield  {author} {\bibinfo {author} {\bibfnamefont {K.}~\bibnamefont
  {Buschow}},\ }\href@noop {} {\bibfield  {journal} {\bibinfo  {journal} {J.
  Magn. Magn. Mater.}\ }\textbf {\bibinfo {volume} {40}},\ \bibinfo {pages}
  {224} (\bibinfo {year} {1983})}\BibitemShut {NoStop}%
\bibitem [{\citenamefont {Parker}\ and\ \citenamefont
  {Oesterreicher}(1983)}]{parker1983magnetic}%
  \BibitemOpen
  \bibfield  {author} {\bibinfo {author} {\bibfnamefont {F.}~\bibnamefont
  {Parker}}\ and\ \bibinfo {author} {\bibfnamefont {H.}~\bibnamefont
  {Oesterreicher}},\ }\href@noop {} {\bibfield  {journal} {\bibinfo  {journal}
  {Journal of the Less Common Metals}\ }\textbf {\bibinfo {volume} {90}},\
  \bibinfo {pages} {127} (\bibinfo {year} {1983})}\BibitemShut {NoStop}%
\bibitem [{\citenamefont {Gottwick}\ \emph {et~al.}(1985)\citenamefont
  {Gottwick}, \citenamefont {Gloss}, \citenamefont {Horn}, \citenamefont
  {Steglich},\ and\ \citenamefont {Grewe}}]{gottwick1985transport}%
  \BibitemOpen
  \bibfield  {author} {\bibinfo {author} {\bibfnamefont {U.}~\bibnamefont
  {Gottwick}}, \bibinfo {author} {\bibfnamefont {K.}~\bibnamefont {Gloss}},
  \bibinfo {author} {\bibfnamefont {S.}~\bibnamefont {Horn}}, \bibinfo {author}
  {\bibfnamefont {F.}~\bibnamefont {Steglich}},\ and\ \bibinfo {author}
  {\bibfnamefont {N.}~\bibnamefont {Grewe}},\ }\href@noop {} {\bibfield
  {journal} {\bibinfo  {journal} {J. Magn. Magn. Mater.}\ }\textbf {\bibinfo
  {volume} {47}},\ \bibinfo {pages} {536} (\bibinfo {year} {1985})}\BibitemShut
  {NoStop}%
\bibitem [{\citenamefont {Tazuke}\ \emph {et~al.}(1993)\citenamefont {Tazuke},
  \citenamefont {Nakabayashi}, \citenamefont {Murayama}, \citenamefont
  {Sakakibara},\ and\ \citenamefont {Goto}}]{tazuke1993magnetism}%
  \BibitemOpen
  \bibfield  {author} {\bibinfo {author} {\bibfnamefont {Y.}~\bibnamefont
  {Tazuke}}, \bibinfo {author} {\bibfnamefont {R.}~\bibnamefont {Nakabayashi}},
  \bibinfo {author} {\bibfnamefont {S.}~\bibnamefont {Murayama}}, \bibinfo
  {author} {\bibfnamefont {T.}~\bibnamefont {Sakakibara}},\ and\ \bibinfo
  {author} {\bibfnamefont {T.}~\bibnamefont {Goto}},\ }\href@noop {} {\bibfield
   {journal} {\bibinfo  {journal} {Phys. B: Condens. Matter}\ }\textbf
  {\bibinfo {volume} {186}},\ \bibinfo {pages} {596} (\bibinfo {year}
  {1993})}\BibitemShut {NoStop}%
\bibitem [{\citenamefont {Tazuke}\ \emph {et~al.}(1997)\citenamefont {Tazuke},
  \citenamefont {Abe},\ and\ \citenamefont {Funahashi}}]{tazuke1997magnetic}%
  \BibitemOpen
  \bibfield  {author} {\bibinfo {author} {\bibfnamefont {Y.}~\bibnamefont
  {Tazuke}}, \bibinfo {author} {\bibfnamefont {M.}~\bibnamefont {Abe}},\ and\
  \bibinfo {author} {\bibfnamefont {S.}~\bibnamefont {Funahashi}},\ }\href@noop
  {} {\bibfield  {journal} {\bibinfo  {journal} {Phys. B: Condens. Matter}\
  }\textbf {\bibinfo {volume} {237}},\ \bibinfo {pages} {559} (\bibinfo {year}
  {1997})}\BibitemShut {NoStop}%
\bibitem [{\citenamefont {Fukase}\ \emph {et~al.}(1999)\citenamefont {Fukase},
  \citenamefont {Tazuke}, \citenamefont {Mitamura}, \citenamefont {Goto},\ and\
  \citenamefont {Sato}}]{fukase1999successive}%
  \BibitemOpen
  \bibfield  {author} {\bibinfo {author} {\bibfnamefont {M.}~\bibnamefont
  {Fukase}}, \bibinfo {author} {\bibfnamefont {Y.}~\bibnamefont {Tazuke}},
  \bibinfo {author} {\bibfnamefont {H.}~\bibnamefont {Mitamura}}, \bibinfo
  {author} {\bibfnamefont {T.}~\bibnamefont {Goto}},\ and\ \bibinfo {author}
  {\bibfnamefont {T.}~\bibnamefont {Sato}},\ }\href@noop {} {\bibfield
  {journal} {\bibinfo  {journal} {J. Phys. Soc. Japan}\ }\textbf {\bibinfo
  {volume} {68}},\ \bibinfo {pages} {1460} (\bibinfo {year}
  {1999})}\BibitemShut {NoStop}%
\bibitem [{\citenamefont {Fukase}\ \emph {et~al.}(2000)\citenamefont {Fukase},
  \citenamefont {Tazuke}, \citenamefont {Mitamura}, \citenamefont {Goto},\ and\
  \citenamefont {Sato}}]{fukase2000itinerant}%
  \BibitemOpen
  \bibfield  {author} {\bibinfo {author} {\bibfnamefont {M.}~\bibnamefont
  {Fukase}}, \bibinfo {author} {\bibfnamefont {Y.}~\bibnamefont {Tazuke}},
  \bibinfo {author} {\bibfnamefont {H.}~\bibnamefont {Mitamura}}, \bibinfo
  {author} {\bibfnamefont {T.}~\bibnamefont {Goto}},\ and\ \bibinfo {author}
  {\bibfnamefont {T.}~\bibnamefont {Sato}},\ }\href@noop {} {\bibfield
  {journal} {\bibinfo  {journal} {Materials transactions, JIM}\ }\textbf
  {\bibinfo {volume} {41}},\ \bibinfo {pages} {1046} (\bibinfo {year}
  {2000})}\BibitemShut {NoStop}%
\bibitem [{\citenamefont {Tazuke}\ \emph {et~al.}(2004)\citenamefont {Tazuke},
  \citenamefont {Suzuki},\ and\ \citenamefont
  {Tanikawa}}]{tazuke2004metamagnetic}%
  \BibitemOpen
  \bibfield  {author} {\bibinfo {author} {\bibfnamefont {Y.}~\bibnamefont
  {Tazuke}}, \bibinfo {author} {\bibfnamefont {H.}~\bibnamefont {Suzuki}},\
  and\ \bibinfo {author} {\bibfnamefont {H.}~\bibnamefont {Tanikawa}},\
  }\href@noop {} {\bibfield  {journal} {\bibinfo  {journal} {Phys. B: Condens.
  Matter}\ }\textbf {\bibinfo {volume} {346}},\ \bibinfo {pages} {122}
  (\bibinfo {year} {2004})}\BibitemShut {NoStop}%
\bibitem [{\citenamefont {Crivello}\ and\ \citenamefont
  {Paul-Boncour}(2020)}]{crivello2020relation}%
  \BibitemOpen
  \bibfield  {author} {\bibinfo {author} {\bibfnamefont {J.-C.}\ \bibnamefont
  {Crivello}}\ and\ \bibinfo {author} {\bibfnamefont {V.}~\bibnamefont
  {Paul-Boncour}},\ }\href@noop {} {\bibfield  {journal} {\bibinfo  {journal}
  {Journal of Physics: Condensed Matter}\ }\textbf {\bibinfo {volume} {32}},\
  \bibinfo {pages} {145802} (\bibinfo {year} {2020})}\BibitemShut {NoStop}%
\bibitem [{\citenamefont {Ribeiro}\ \emph {et~al.}(2022)\citenamefont
  {Ribeiro}, \citenamefont {Bud'ko}, \citenamefont {Xiang}, \citenamefont
  {Ryan},\ and\ \citenamefont {Canfield}}]{ribeiro2022small}%
  \BibitemOpen
  \bibfield  {author} {\bibinfo {author} {\bibfnamefont {R.~A.}\ \bibnamefont
  {Ribeiro}}, \bibinfo {author} {\bibfnamefont {S.~L.}\ \bibnamefont {Bud'ko}},
  \bibinfo {author} {\bibfnamefont {L.}~\bibnamefont {Xiang}}, \bibinfo
  {author} {\bibfnamefont {D.~H.}\ \bibnamefont {Ryan}},\ and\ \bibinfo
  {author} {\bibfnamefont {P.~C.}\ \bibnamefont {Canfield}},\ }\href@noop {}
  {\bibfield  {journal} {\bibinfo  {journal} {Phys. Rev. B}\ }\textbf {\bibinfo
  {volume} {105}},\ \bibinfo {pages} {014412} (\bibinfo {year}
  {2022})}\BibitemShut {NoStop}%
\bibitem [{\citenamefont {Momma}\ and\ \citenamefont
  {Izumi}(2011)}]{Momma_2011}%
  \BibitemOpen
  \bibfield  {author} {\bibinfo {author} {\bibfnamefont {K.}~\bibnamefont
  {Momma}}\ and\ \bibinfo {author} {\bibfnamefont {F.}~\bibnamefont {Izumi}},\
  }\href@noop {} {\bibfield  {journal} {\bibinfo  {journal} {J. Appl.
  Crystallogr.}\ }\textbf {\bibinfo {volume} {44}},\ \bibinfo {pages} {1272}
  (\bibinfo {year} {2011})}\BibitemShut {NoStop}%
\bibitem [{\citenamefont {Canfield}\ and\ \citenamefont
  {Fisher}(2001)}]{canfield2001high}%
  \BibitemOpen
  \bibfield  {author} {\bibinfo {author} {\bibfnamefont {P.~C.}\ \bibnamefont
  {Canfield}}\ and\ \bibinfo {author} {\bibfnamefont {I.~R.}\ \bibnamefont
  {Fisher}},\ }\href@noop {} {\bibfield  {journal} {\bibinfo  {journal}
  {Journal of Crystal Growth}\ }\textbf {\bibinfo {volume} {225}},\ \bibinfo
  {pages} {155} (\bibinfo {year} {2001})}\BibitemShut {NoStop}%
\bibitem [{\citenamefont {Di}\ \emph {et~al.}(2000)\citenamefont {Di},
  \citenamefont {Yamamoto}, \citenamefont {Inui},\ and\ \citenamefont
  {Yamaguchi}}]{di2000characterization}%
  \BibitemOpen
  \bibfield  {author} {\bibinfo {author} {\bibfnamefont {Z.}~\bibnamefont
  {Di}}, \bibinfo {author} {\bibfnamefont {T.}~\bibnamefont {Yamamoto}},
  \bibinfo {author} {\bibfnamefont {H.}~\bibnamefont {Inui}},\ and\ \bibinfo
  {author} {\bibfnamefont {M.}~\bibnamefont {Yamaguchi}},\ }\href@noop {}
  {\bibfield  {journal} {\bibinfo  {journal} {Intermetallics}\ }\textbf
  {\bibinfo {volume} {8}},\ \bibinfo {pages} {391} (\bibinfo {year}
  {2000})}\BibitemShut {NoStop}%
\bibitem [{\citenamefont {Aroyo}\ \emph {et~al.}(2006)\citenamefont {Aroyo},
  \citenamefont {Perez-Mato}, \citenamefont {Capillas}, \citenamefont
  {Kroumova}, \citenamefont {Ivantchev}, \citenamefont {Madariaga},
  \citenamefont {Kirov},\ and\ \citenamefont {Wondratschek}}]{aroyo2006bilbao}%
  \BibitemOpen
  \bibfield  {author} {\bibinfo {author} {\bibfnamefont {M.~I.}\ \bibnamefont
  {Aroyo}}, \bibinfo {author} {\bibfnamefont {J.~M.}\ \bibnamefont
  {Perez-Mato}}, \bibinfo {author} {\bibfnamefont {C.}~\bibnamefont
  {Capillas}}, \bibinfo {author} {\bibfnamefont {E.}~\bibnamefont {Kroumova}},
  \bibinfo {author} {\bibfnamefont {S.}~\bibnamefont {Ivantchev}}, \bibinfo
  {author} {\bibfnamefont {G.}~\bibnamefont {Madariaga}}, \bibinfo {author}
  {\bibfnamefont {A.}~\bibnamefont {Kirov}},\ and\ \bibinfo {author}
  {\bibfnamefont {H.}~\bibnamefont {Wondratschek}},\ }\href@noop {} {\bibfield
  {journal} {\bibinfo  {journal} {Zeitschrift f{\"u}r
  Kristallographie-Crystalline Materials}\ }\textbf {\bibinfo {volume} {221}},\
  \bibinfo {pages} {15} (\bibinfo {year} {2006})}\BibitemShut {NoStop}%
\bibitem [{\citenamefont {Rodriguez-Carvajal}(1993)}]{rodriguez1993fullprof}%
  \BibitemOpen
  \bibfield  {author} {\bibinfo {author} {\bibfnamefont {J.}~\bibnamefont
  {Rodriguez-Carvajal}},\ }\href@noop {} {\bibfield  {journal} {\bibinfo
  {journal} {Physica B}\ }\textbf {\bibinfo {volume} {192}},\ \bibinfo {pages}
  {55} (\bibinfo {year} {1993})}\BibitemShut {NoStop}%
\bibitem [{\citenamefont {Werwi{\'n}ski}\ \emph {et~al.}(2018)\citenamefont
  {Werwi{\'n}ski}, \citenamefont {Szajek}, \citenamefont {Marczy{\'n}ska},
  \citenamefont {Smardz}, \citenamefont {Nowak},\ and\ \citenamefont
  {Jurczyk}}]{werwinski2018effect}%
  \BibitemOpen
  \bibfield  {author} {\bibinfo {author} {\bibfnamefont {M.}~\bibnamefont
  {Werwi{\'n}ski}}, \bibinfo {author} {\bibfnamefont {A.}~\bibnamefont
  {Szajek}}, \bibinfo {author} {\bibfnamefont {A.}~\bibnamefont
  {Marczy{\'n}ska}}, \bibinfo {author} {\bibfnamefont {L.}~\bibnamefont
  {Smardz}}, \bibinfo {author} {\bibfnamefont {M.}~\bibnamefont {Nowak}},\ and\
  \bibinfo {author} {\bibfnamefont {M.}~\bibnamefont {Jurczyk}},\ }\href@noop
  {} {\bibfield  {journal} {\bibinfo  {journal} {Journal of Alloys and
  Compounds}\ }\textbf {\bibinfo {volume} {763}},\ \bibinfo {pages} {951}
  (\bibinfo {year} {2018})}\BibitemShut {NoStop}%
\bibitem [{\citenamefont {Singh}(2015)}]{singh2015electronic}%
  \BibitemOpen
  \bibfield  {author} {\bibinfo {author} {\bibfnamefont {D.~J.}\ \bibnamefont
  {Singh}},\ }\href@noop {} {\bibfield  {journal} {\bibinfo  {journal} {Phys.
  Rev. B}\ }\textbf {\bibinfo {volume} {92}},\ \bibinfo {pages} {174403}
  (\bibinfo {year} {2015})}\BibitemShut {NoStop}%
\bibitem [{\citenamefont {Werwi{\'n}ski}\ \emph {et~al.}(2019)\citenamefont
  {Werwi{\'n}ski}, \citenamefont {Szajek}, \citenamefont {Marczy{\'n}ska},
  \citenamefont {Smardz}, \citenamefont {Nowak},\ and\ \citenamefont
  {Jurczyk}}]{werwinski2019effect}%
  \BibitemOpen
  \bibfield  {author} {\bibinfo {author} {\bibfnamefont {M.}~\bibnamefont
  {Werwi{\'n}ski}}, \bibinfo {author} {\bibfnamefont {A.}~\bibnamefont
  {Szajek}}, \bibinfo {author} {\bibfnamefont {A.}~\bibnamefont
  {Marczy{\'n}ska}}, \bibinfo {author} {\bibfnamefont {L.}~\bibnamefont
  {Smardz}}, \bibinfo {author} {\bibfnamefont {M.}~\bibnamefont {Nowak}},\ and\
  \bibinfo {author} {\bibfnamefont {M.}~\bibnamefont {Jurczyk}},\ }\href@noop
  {} {\bibfield  {journal} {\bibinfo  {journal} {Journal of Alloys and
  Compounds}\ }\textbf {\bibinfo {volume} {773}},\ \bibinfo {pages} {131}
  (\bibinfo {year} {2019})}\BibitemShut {NoStop}%
\bibitem [{\citenamefont {Lee}\ \emph {et~al.}(2022)\citenamefont {Lee},
  \citenamefont {Jo}, \citenamefont {Wang}, \citenamefont {Ribeiro},
  \citenamefont {Kushnirenko}, \citenamefont {Schrunk}, \citenamefont
  {Canfield},\ and\ \citenamefont {Kaminski}}]{lee2022electronic}%
  \BibitemOpen
  \bibfield  {author} {\bibinfo {author} {\bibfnamefont {K.}~\bibnamefont
  {Lee}}, \bibinfo {author} {\bibfnamefont {N.~H.}\ \bibnamefont {Jo}},
  \bibinfo {author} {\bibfnamefont {L.-L.}\ \bibnamefont {Wang}}, \bibinfo
  {author} {\bibfnamefont {R.}~\bibnamefont {Ribeiro}}, \bibinfo {author}
  {\bibfnamefont {Y.}~\bibnamefont {Kushnirenko}}, \bibinfo {author}
  {\bibfnamefont {B.}~\bibnamefont {Schrunk}}, \bibinfo {author} {\bibfnamefont
  {P.~C.}\ \bibnamefont {Canfield}},\ and\ \bibinfo {author} {\bibfnamefont
  {A.}~\bibnamefont {Kaminski}},\ }\href@noop {} {\bibfield  {journal}
  {\bibinfo  {journal} {arXiv preprint arXiv:2205.09211}\ } (\bibinfo {year}
  {2022})}\BibitemShut {NoStop}%
\end{thebibliography}%

\end{document}